\documentclass[twocolumn,tighten,times]{aastex63}
\usepackage{graphicx}
\usepackage{amsmath}
\usepackage{natbib}
\usepackage{amssymb}

\usepackage{sidecap}

\shorttitle{Cataclysmic Deaths of CVs}
\shortauthors{Metzger et al.}

\begin{document}

\newcommand{\be}{\begin{equation}}
\newcommand{\ee}{\end{equation}}

\title{Transients from the Cataclysmic Deaths of Cataclysmic Variables}

\author[0000-0002-4670-7509]{Brian D. Metzger}
\affil{Department of Physics and Columbia Astrophysics Laboratory, Columbia University, Pupin Hall, New York, NY 10027, USA}
\affil{Center for Computational Astrophysics, Flatiron Institute, 162 5th Ave, New York, NY 10010, USA} 

\author[0000-0002-0632-8897]{Yossef Zenati}
\affil{Department of Physics and Astronomy, Johns Hopkins University, Baltimore, MD 21218, USA}
\affil{CHE Israel Excellence Fellowship}

\author[0000-0002-8400-3705]{Laura Chomiuk}
\affil{Center for Data Intensive and Time Domain Astronomy, Department of Physics and Astronomy, Michigan State University, East Lansing, MI 48824, USA}

\author[0000-0002-9632-6106]{Ken J.~Shen}
\affil{Department of Astronomy and Theoretical Astrophysics Center, University of California, Berkeley, CA 94720, USA}

\author[0000-0002-1468-9668]{Jay Strader}
\affil{Center for Data Intensive and Time Domain Astronomy, Department of Physics and Astronomy, Michigan State University, East Lansing, MI 48824, USA}

\begin{abstract}
We explore the observational appearance of the merger of a low-mass star with a white dwarf (WD) binary companion.  We are motivated by recent work finding that multiple tensions between the observed properties of cataclysmic variables (CVs) and standard evolution models are resolved if a large fraction of CV binaries merge as a result of unstable mass transfer.  Tidal disruption of the secondary forms a geometrically thick disk around the WD, which subsequently accretes at highly super-Eddington rates.  Analytic estimates and numerical hydrodynamical simulations reveal that outflows from the accretion flow unbind a large fraction $\gtrsim 90\%$ of the secondary at velocities $\sim 500-1000$ km s$^{-1}$ within days of the merger.  Hydrogen recombination in the expanding ejecta powers optical transient emission lasting about a month with a luminosity $\gtrsim 10^{38}$ erg s$^{-1}$, similar to slow classical novae and luminous red novae (LRN) from ordinary stellar mergers.  Over longer timescales the mass accreted by the WD undergoes hydrogen shell burning, inflating the remnant into a giant of luminosity $\sim 300-5000L_{\odot}$, effective temperature $T_{\rm eff} \approx 3000$ K and lifetime $\sim 10^{4}-10^{5}$ yr.  We predict that $\sim 10^{3}-10^{4}$ Milky Way giants are CV merger products, potentially distinguishable by atypical surface abundances.  We explore whether any Galactic historical slow classical novae are masquerading CV mergers by identifying four such post-nova systems with potential giant counterparts for which a CV merger origin cannot be ruled out.  We address whether the historical transient CK Vul and its gaseous/dusty nebula resulted from a CV merger.
\end{abstract}


\section{Introduction}
\label{sec:introduction}

Cataclysmic variables (CVs) are semi-detached binaries in which a main sequence or moderately evolved hydrogen-rich star transfers mass onto a white dwarf (WD) primary (e.g., \citealt{Patterson84,Kolb93,Warner95}).  CVs provide key laboratories for studying the physics of binary mass transfer (e.g., \citealt{King+95}), nucleosynthesis (e.g., \citealt{Jose+06}), disk accretion (e.g., \citealt{Dubus+18}) and even jet formation (e.g., \citealt{Coppejans&Knigge20}).  The standard model of CV evolution postulates that the binary properties over time are driven primarily by angular momentum loss due to a magnetized wind from the secondary and gravitational wave radiation (e.g., \citealt{Rappaport+83,Spruit&Ritter83}; however, see \citealt{Knigge+00,Ginzburg&Quataert21}).  A wide range of observational evidence supports the general features of this scenario (e.g., \citealt{Townsley&Bildsten03,Knigge06,Schreiber+10}).

Despite these successes, a number of nagging discrepancies have long persisted between CV observations and population modeling.  The space density of CVs is found to be $\sim 10-100$ times lower than theoretically predicted (e.g., \citealt{deKool92,Patterson98,Schreiber&Gansicke03,Pretorius&Knigge12,Pala+20}) and the minimum CV orbital period is longer than expected (e.g., \citealt{Gansicke+09,Knigge+11}).  Perhaps most puzzling, the WD masses in CVs are systematically larger than those in their progenitor population, the post common-envelope detached binaries (e.g., \citealt{Gansicke+09,Zorotovic+11}).  The formation of a large number of CVs containing low-mass white dwarfs appears to be a generic consequence of CV models, regardless of the assumptions (e.g., \citealt{deKool92,Kolb93,Politano96,Zorotovic+11}).  \citet{Schreiber+16} and \citet{Belloni+18} find that all of these tensions are alleviated if CV binaries experience an additional sink of angular momentum beyond that due to magnetic braking and gravitational waves, an unidentified empirically motivated source of ``consequential angular momentum loss" (CAML; see \citealt{Zorotovic&Schreiber20} for a review).\footnote{The mechanism is ``consequential" insofar the angular momentum loss is a direct result of the mass transfer process, unlike gravitational radiation or stellar winds.}  In effect, CAML causes a significant fraction of CVs to ``drop out" of the population as a result of the binary being destroyed in a merger following the onset of unstable mass transfer.  However, in order to explain the observed CV population, the CAML mechanism must preferentially act on CVs with lower mass WDs.

A promising mechanism for the CAML identified by \citet{Schreiber+16} are classical novae, i.e. thermonuclear outbursts that occur on the WD surface due to unstable hydrogen burning (\citealt{Gallagher&Starrfield76,Chomiuk+20}).  Following the thermonuclear runaway, the outer WD envelope expands to encompass the secondary star, resulting in a physical situation somewhat similar to the ``common envelope" interaction between non-degenerate binary stars (e.g., \citealt{Ivanova+13}).  Gas drag on the binary can reduce its angular momentum \citep{Macdonald+85,Shankar+91,Livio+91,Schenker+98}, potentially destabilizing the system and leading to a merger \citep{Shen15,Nelemans+16,Chomiuk+20}.\footnote{There may be other ways by which novae destroy their companion stars, in a comparatively rapid but non-dynamical manner.  Some CVs exhibit elevated mass transfer rates triggered by irradiation of the secondary by the nova outburst (e.g., \citealt{Ginzburg&Quataert21}); insofar as novae occur more frequently for higher mass transfer rates, this could in principle a positive feedback cycle that erodes the companion mass within millions of years (e.g., \citealt{Knigge+00,Patterson+13}).}  Indeed, the higher envelopes masses and longer duration of novae expected to take place on lower-mass WDs would render frictional drag particularly effective in these systems (e.g., \citealt{Kato&Hachisu11,Liu&Li19}).  Other processes during novae could in principle also lead to  angular momentum loss, such as the “braking” interaction between the secondary's magnetic field and the nova ejecta \citep{Martin+11}, asymmetric expulsion of nova ejecta (\citealt{Nelemans+16,Schaefer+19}), or torques from a circumbinary disk (\citealt{Taam&Spruit01,Liu&Li16}).\footnote{By contrast, if angular momentum is largely conserved during a novae, then the binary separation will expand; this may lead to a decline in accretion rate compared to that just before the novae, and is the origin of the hypothesis that CVs “hibernate” after a nova eruption \citep{Prialnik&Shara86,Shara+86,Kovetz+88,Hillman+20}.}  By comparing the orbital periods of CVs before and after a nova eruption, \citet{Schaefer20b} find evidence that at least some novae remove angular momentum from the binary.

In this paper we explore the direct observational signatures of the ``deaths'' of CVs in dynamical mergers.  The final outcome of the unstable mass transfer process is the tidal disruption of the secondary star, resulting in the formation of a massive hydrogen-rich disk around the WD.  As we shall describe, the subsequent accretion of this disk onto the WD occurs on a timescale as short as days at super-Eddington rates.  Such super-Eddington accretion flows are subject to powerful outflows from the disk  which eject a majority of the secondary's mass.  As these wind ejecta expand into space and become transparent, they power a $\sim$weeks to months long optical transient, with light curve properties broadly similar to those of slow classical novae and the ``luminous red novae'' (LRN) which accompany the mergers of two ordinary (i.e., non-degenerate) stars (e.g., \citealt{Bond+03,Soker&Tylenda06,Tylenda+11}).  However, unlike ordinary stellar mergers in which the final remnant is typically an ordinary non-degenerate star (albeit one out of thermal equilibrium; e.g., \citealt{Hoadley+20}), the final remnant of a CV merger is WD with a hydrogen burning shell and a luminosity greatly exceeding that of the original CV.

The historical transient CK Vulpeculae (Nova Vulpeculae 1670), long thought to be a nova, has in recent years been argued to be a stellar merger \citep{Kaminski+15,Kaminski+20a,Kaminski+20b} or a merger between a brown dwarf and a WD (\citealt{Eyres+18}).  Another motivation for our work is thus to explore whether CK Vul is consistent with being a CV merger, or a related kind of event involving the merger of an eccentric white dwarf-star merger in a triple system.

This paper is organized as follows.  In Section \ref{sec:analytic} we provide analytic estimates of the properties of the disk formed during the merger and its mass outflows.  In Section \ref{sec:numerical} we present axisymmetric hydrodynamical simulations of the post-merger disk evolution which quantify the properties of the accreted matter and disk outflows.  In Section \ref{sec:transient} we discuss the immediate and long-term transient signatures of CV mergers.  In Section \ref{sec:discussion} we discuss several implications of our results and perform a systematic analysis of historic Galactic slow novae to determine if any of their remnants are consistent with being CV merger products (Section \ref{sec:slownovae}; Appendix \ref{sec:candidates}).  In Section \ref{sec:conclusion} we summarize our findings and conclude.

\section{Disk Formation and Outflows}
\label{sec:analytic}

We begin by providing analytic estimates of the immediate outcome of the merger, which provide insight into the transient WD accretion phase, disk outflows, and the key timescales involved.  These considerations also motivate the initial conditions for our numerical simulations in Section \ref{sec:numerical}.

\subsection{Initial Binary Properties and Disk Formation}

We are interested in the fate of unstable mass transfer in a binary system consisting of a WD primary of mass $M_{\rm WD}$ and radius $R_{\rm WD}$ orbited by a secondary companion of mass $M_{\star} \lesssim M_{\rm WD}$ and radius $R_{\star} \gg R_{\rm WD}$. The companion star is nominally a low-mass main sequence star or brown dwarf, as characterizes CVs before and after the period minimum, respectively.  \citet{Zorotovic&Schreiber17} find secondary masses at the point of disruption that span a wide range from $\sim 0.05 M_{\odot}$ to $\gtrsim 0.6M_{\odot}$.

Mass transfer occurs as the binary loses orbital angular momentum, resulting in Roche-lobe overflow (RLOF) of the secondary onto the primary.  For circular orbits, this takes place at an orbital separation \citep{Eggleton83}
\be
a_{\rm RLOF} \approx R_{\star}\frac{0.6 q^{2/3}+ {\rm ln}(1+q^{1/3})}{0.49 q^{2/3}} \underset{q \ll 1}\approx 2.16 q^{-1/3} R_{\star},
\label{eq:RLOF}
\ee
where $q \equiv M_{\star}/M_{\rm WD}$.  In the final equality we have taken the limit $q \ll 1$, in which case this expression essentially reduces to the orbital semi-major axis $a$ at which the stellar radius equals the Hill's sphere radius, $R_{\rm H} \approx a(M_{\star}/M_{\rm WD})^{1/3}$.  For the mass-radius relationship of the companion, spanning the massive planet to low-mass star range, we take (e.g., \citealt{Chabrier+09})
\begin{equation} R_{\star} \approx \begin{cases}
R_{\odot}\left(\frac{M_{\star}}{M_{\odot}}\right)^{0.8} & 0.1 \lesssim M_{\star}/M_{\odot} \lesssim 1  \\
0.1R_{\odot} & 10^{-3} \lesssim M_{\star}/M_{\odot} \lesssim 0.1,
\end{cases}
\label{eq:Rstar}
\end{equation}
For the radius of the WD \citep{Nauenberg72},
\be
R_{\rm WD} \approx 10^{9}\,{\rm cm}\left(\frac{M_{\rm WD}}{0.7M_{\odot}}\right)^{-1/3}\left[1 - \left(\frac{M_{\rm WD}}{M_{\rm ch}}\right)^{4/3}\right]^{1/2},
\ee
where $M_{\rm ch} \approx 1.45M_{\odot}$.

The process of unstable mass transfer leads to a runaway increase in the mass transfer rate and, ultimately, the tidal disruption of the companion star by the WD.  At the end of this process, the companion is quickly$-$on a few orbital periods$-$sheared into an accretion disk (e.g., as illustrated by numerical simulations of other unstable mass-transfer events, such as mergers between WDs and black holes; e.g., \citealt{Fryer+99}).  The characteristic radial dimension of the disk can be estimated as (e.g.,~\citealt{Margalit&Metzger16})
\be
R_{\rm d,0} = a_{\rm RLOF}(1 + q)^{-1}.
\label{eq:Rc}
\ee
This is the semi-major axis of a point mass $\sim M_{\star}$ in orbit around the WD, with an angular momentum equal to that of the binary at the time of disruption (which is assumed to be conserved during the disruption process).  

The mass of the formed disk will likewise be approximately equal to that of the original secondary, $M_{\rm d,0} \approx M_{\star}$.  However, we note that even prior to the dynamical merger phase, appreciable mass may be lost from the system (e.g., through the outer, $L_{2}$ Lagrange point; \citealt{Pejcha+17}) in which case $M_{\rm d,0}$ will be somewhat smaller than $M_{\star}$; in Section \ref{sec:transient} we discuss the impact of pre-dynamical mass loss on the merger's transient emission.
 
\subsection{Initial Disk Properties}

We now estimate the properties of the disk created from the disrupted secondary, immediately after its formation (an epoch we denote by the subscript `0').  We work in the $q \ll 1$ limit so the results can be readily scaled from planets ($q \sim 10^{-3}$) to brown dwarfs ($q \sim 0.1$).  This limit also gives quantitatively reasonable results for low-mass stars ($q \sim 1$).

Combining Eqs.~(\ref{eq:RLOF}) and (\ref{eq:Rc}), we obtain the initial outer radius of the disk,
\be
R_{\rm d,0} \approx 0.5 q_{0.1}^{-1/3}R_{0.1}R_{\odot},
\ee
where $q_{0.1} \equiv q/(0.1)$, $R_{0.1} \equiv R_{\star}/(0.1R_{\odot})$.  We note that $R_{\rm d,0}$ is typically $\sim 100R_{\rm WD}$.

Assuming an initial disk mass $M_{\rm d,0} \sim M_{\star}$ (i.e., neglecting pre-dynamical mass loss), the characteristic initial surface density of the disk is 
\be
\Sigma_0 \sim \frac{M_{\rm d,0}}{2\pi R_{\rm d, 0}^{2}} \approx 1.6\times 10^{10}\,{\rm g\,cm^{-2}}q_{0.1}^{5/3}M_{0.6}R_{0.1}^{-2},
\ee
where $M_{0.6} \equiv M_{\rm WD}/(0.6M_{\odot})$.  Due to the gravitational energy released by the disruption process, and the inability to cool efficiently (see below), the initial disk will be hot and geometrically thick after forming, with a vertical scale-height $H_0$ and aspect ratio $\theta_0 \equiv H_0/R_{\rm d,0} \sim 1/3$ \citep{Metzger12,Margalit&Metzger16}.  The density in the disk midplane at $r \sim R_{\rm d,0}$ is then
\be
\rho_0 \simeq \frac{\Sigma}{2H_0} \approx 0.7\,{\rm g\,cm^{-3}}q_{0.1}^{2}M_{0.6}R_{0.1}^{-2}\theta_{0.33}^{-1},
\label{eq:rho0}
\ee
where $\theta_{0.33} \equiv \theta_0/(0.33)$.  Under the assumption that ideal gas pressure dominates, the midplane temperature of the disk at $r \sim R_{\rm d,0}$ is
\be
T_{\rm 0} \simeq \frac{GM_{\rm WD}\,\mu m_p}{k R_{\rm d,0}}\theta^{2} \approx 1.7\times 10^{6}\,{\rm K}\,\,q_{0.1}^{1/3}M_{0.6}R_{0.1}^{-1}\theta_{0.33}^{2}
\label{eq:T0}
\ee
where $\mu \simeq 0.62$ is the mean molecular weight of fully ionized solar composition material. 

The assumption that gas pressure dominates can be justified by a comparison to other sources of pressure, such as radiation pressure $P_{\rm rad}$ and degeneracy pressure $P_{\rm deg}$.  In particular, using Eqs.~(\ref{eq:rho0}) and (\ref{eq:T0}) we find,
\be
\left.\frac{P_{\rm rad}}{P_{\rm gas}}\right|_{R_{\rm d,0}} = \frac{a \mu m_p T_{0}^{3}}{3 \rho_0 k} \sim 2\times 10^{-4}M_{0.6}^{2}R_{0.1}\theta_{0.33}^{7}q_{0.1}^{-1}
\label{eq:Pratio}
\ee
\begin{eqnarray}
\left.\frac{P_{\rm deg}}{P_{\rm gas}}\right|_{R_{\rm d,0}} &\approx& \frac{h^{2}}{20 m_e k T_{0}}\left(\frac{3}{\pi}\right)^{2/3}\left(\frac{\mu}{\mu_e}\right)\left(\frac{\rho_0}{\mu m_p}\right)^{2/3} \nonumber \\
&\sim& 0.07 q_{0.1}M_{0.6}^{-1/3}R_{0.1}^{-1/3}\theta_{0.33}^{-8/3},
\end{eqnarray}
where $\mu_e \simeq 1.3$ is the mean molecular weight per electron.  Thus, both radiation and degeneracy pressure are typically subdominant to gas pressure at radii $\sim R_{\rm d,0}$, though radiation pressure can become important close to the WD surface.

\subsection{Accretion Phase}

After forming, the disk will begin to accrete onto the WD as a result of angular momentum transport driven by the magnetorotational instability (MRI; \citealt{Balbus&Hawley98}), and possibly by gravitational instabilities (see below).  The ``viscous'' timescale, over which the peak accretion rate is reached, can be estimated as (e.g., \citealt{Frank+02})
\begin{eqnarray}
t_{\rm visc,0} &\sim& \left.\frac{r^{2}}{\nu}\right|_{R_{\rm d,0}} \sim \frac{1}{\alpha}\frac{1}{\theta_0^{2}}\left(\frac{R_{\rm d,0}^{3}}{GM_{\rm WD}}\right)^{1/2} \nonumber \\
&\sim& 7\times 10^{4}\,{\rm s}\, \alpha_{0.1}^{-1}q_{0.1}^{-1/2}R_{0.1}^{3/2}M_{0.6}^{-1/2}\theta_{0.33}^{-2},  \label{eq:tvisc}
\end{eqnarray}
where $\nu = \alpha c_{\rm s}H = \alpha r^{2}\Omega_{\rm K}\theta^{2}$ is the effective kinematic viscosity, $\Omega_{\rm K} \equiv (GM_{\star}/r^{3})^{1/2}$ is the Keplerian orbital frequency, $c_{\rm s} \approx H\Omega_{\rm K}$ is the midplane sound speed, and $\alpha = 0.1\alpha_{0.1}$ is the viscosity parameter \citep{Shakura&Sunyaev73} scaled to a typical value (e.g., \citealt{King+07}).  The viscous timescale typically ranges from a day to a week.

On timescales $t \gtrsim t_{\rm visc,0}$, the disk will establish a steady flow onto the WD surface (e.g., \citealt{Frank+02}).  The characteristic peak accretion rate is approximately,
\be
\dot{M}_0 \sim \frac{M_{\rm d,0}}{t_{\rm visc,0}} \sim 2\times 10^{27}{\rm g\,s^{-1}}\,\alpha_{0.1}q_{0.1}^{3/2}R_{0.1}^{-3/2}M_{0.6}^{3/2}\theta_{0.33}^{2}.
\ee
This is typically $\sim 4-5$ orders of magnitude larger than the WD Eddington accretion rate $\dot{M}_{\rm Edd} \sim 10^{21}$ g s$^{-1}$, justifying our earlier assumption of a geometrically thick disk.  

When $\dot{M} \gg \dot{M}_{\rm Edd}(R_{\rm 0}/R_{\rm WD}) \sim 100 \dot{M}_{\rm Edd}$, photons are trapped and advected inwards in the disk at radii $\lesssim R_{\rm d,0}$ and hence the disk cannot cool efficiently through radiation (e.g., \citealt{Shakura&Sunyaev73}).  In this situation, the accretion flow is susceptible to significant mass outflows powered by the released gravitational energy (e.g., \citealt{Narayan&Yi95,Blandford&Begelman99,Kitaki+21}).  As a result of outflows, the mass inflow rate decreases approaching the WD surface, in a way typically parametrized by a power-law in radius, viz.~
\be
\dot{M}(r) = \dot{M}_0 \left(\frac{r}{R_{\rm d,0}}\right)^{p},
\label{eq:Mdotr}
\ee
where the value $0 \lesssim p \lesssim 1$ \citep{Blandford&Begelman99} depends on the mechanism driving outflows from the disk.  In what follows, we take $p = 0.6$, motivated by hydrodynamical simulations of radiatively inefficient accretion flows (e.g., \citealt{Yuan&Narayan14}). 
Appendix \ref{sec:1D} presents a one-dimensional (height-integrated) model of the steady-state radial disk structure (i.e., as achieved on the timescale $\sim t_{\rm visc,0}$).

For mass ratios $q \gtrsim 0.1$ of interest to CV mergers, the disk could be sufficiently massive to experience instabilities arising from self-gravity.  This occurs for values of the \citet{Toomre64} parameter,
\be
Q = \frac{\Omega c_s}{\pi G \Sigma} \approx \frac{\Omega_{\rm K}^{2}}{2\pi G \rho} \simeq \frac{M_{\rm WD}(1+q)}{\pi r^{2}\Sigma}\theta \underset{r = R_{\rm d,0}}\sim \frac{1+q}{q}\theta,
\ee
less than a critical value $Q_0 \sim \mathcal{O}(1)$.  For example, taking $\theta_0 = 1/3$ and $Q_0 = 1-1.4$, gravitational instabilities set in for $M_{\star}/M_{\rm WD} \gtrsim 0.3-0.5$.  Their likely effect is to generate non-axisymmetric structures in the disk, such as spiral density waves, which mediate rapid angular momentum transport, reducing the disk mass to the point of marginal stability $Q \approx Q_0$ (e.g., \citealt{Laughlin&Bodenheimer94,Gammie01}).  A ``burst'' of accretion may thus be expected immediately following the disruption, followed by more gradual accretion on the timescale $t_{\rm visc}$ due to the MRI (Eq.~\ref{eq:tvisc}).  
At late times $t \gg t_{\rm visc,0}$, the outer edge of the disk will continue to spread outwards due to the redistribution of angular momentum, its radius growing as
\be
R_{\rm d} \sim R_{\rm d,0}\left(\frac{t}{t_{\rm visc,0}}\right)^{2/3}, t \gg t_{\rm visc,0}
\ee
The accretion rate at $r < R_{\rm d}$ will likewise drop as a power-law (e.g., \citealt{Metzger+08}), viz.~
\be
\dot{M} \propto r^{p}t^{-4(p+1)/3} \propto t^{-2.1}, t \gg t_{\rm visc,0},
\ee
where in the final line we again take $p = 0.6$.  

\subsection{Mass Outflows}

As discussed above, significant mass outflows will occur from the disk on a timescale $\sim t_{\rm visc,0}$.  The total wind mass-loss rate,
\be
\dot{M}_{\rm w} = \dot{M}(R_{\rm d,0})-\dot{M}(R_{\rm WD}) = \dot{M}_{0}\left[1-\left(\frac{R_{\rm WD}}{R_{\rm d,0}}\right)^{p}\right],
\label{eq:Mdotw}
\ee 
is comparable to the total inflow rate $\sim \dot{M}_0$ because $R_{\rm d,0} \gg R_{\rm WD}$.  As a result, most of the companion's mass will be unbound, with only a small fraction accreted onto the WD surface,
\be
\frac{M_{\rm acc}}{M_{\star}} \approx \frac{\dot{M}(R_{\rm WD})}{\dot{M}(R_{\rm d,0})} \sim  \left(\frac{R_{\rm WD}}{R_{\rm d,0}}\right)^{p} \underset{q \ll 1}\approx 0.12\, q_{0.1}^{0.2}R_{0.1}^{-0.6}.
\label{eq:Macc}
\ee
This expression follows from Eq.~(\ref{eq:Mdotr}) in the limit $R_{\rm d,0} \gg R_{\rm WD}$, where in the numerical evaluation we again take $p = 0.6$ and $R_{\rm WD} \approx 10^{9}$ cm.  

For the same value of $p = 0.6$, \citet{Margalit&Metzger16} estimate that the disk outflows will achieve an asymptotic velocity $v_{\rm w} \approx 1.2 v_{\rm K}$, where $v_{\rm K} = r \Omega_{\rm K}$ is the Keplerian orbital speed (see their Fig.~3).  The bulk of the wind will thus emerge from radii $\sim R_{\rm d,0}$ with a velocity (Eq.~\ref{eq:vw})
\begin{eqnarray}
\langle v_{\rm w} \rangle &\sim& 1.2 v_{\rm K}|_{R_{\rm d,0}} \approx 1.2\left(\frac{GM_{\rm WD}}{R_{\rm d,0}}\right)^{1/2} \nonumber \\
&\approx& 570\,{\rm km\,s^{-1}}\, M_{0.6}^{1/2}q_{0.1}^{1/6}R_{0.1}^{-1/2},
\label{eq:vw}
\end{eqnarray}
carrying a kinetic energy
\be
E_{\rm k} \sim \frac{1}{2}M_{\star}\langle v_{\rm w} \rangle^{2} \approx 2\times 10^{47}{\rm erg}\,M_{0.6}^{2}q_{0.1}^{4/3}R_{0.1}^{-1}.
\label{eq:Ek}
\ee
Outflows from the innermost regions of the disk can in principle reach much higher velocities, closer to the escape speed near the WD surface,
\be
v_{\rm max} \sim \left(\frac{GM_{\rm WD}}{2R_{\rm WD}}\right)^{1/2} \approx 2000\,{\rm km\,s^{-1}}.
\label{eq:vfast}
\ee
However, such outflows are short-lived because the inner disk is rapidly truncated by the hot envelope that forms on the WD surface from the accreted material.  

\subsection{Nuclear burning in the disk?}
\label{sec:burning}

In principle, nuclear burning could have a large dynamical effect on the disk evolution, as occurs in the merger of a WD with a neutron star or black hole (e.g., \citealt{Metzger12,Fernandez+2013,Zenati19b,Fernandez+19,Zenati+20,Bobrick+21}).  This is because the energy available through hydrogen burning, $Q \approx 6.3$ MeV per nucleon, greatly exceeds the gravitational binding energy of the disk $\lesssim GM_{\rm WD}m_p/R_{\rm WD} \sim 0.1$ MeV per nucleon.  However, this energy will only be released if the inflowing matter has enough time to burn.   

Nuclear burning will be important if the burning timescale, $t_{\rm nuc}$, at a given radius in the disk midplane is shorter than the local viscous (radial inflow) timescale,
\be
t_{\rm visc} = \frac{r^{2}}{\nu} \sim 3.2\times 10^{2}\,{\rm s}\,\alpha_{0.1}^{-1}M_{0.6}^{-1}\theta_{0.33}^{-2}\left(\frac{r}{R_{\rm WD}}\right)^{3/2},
\label{eq:tvisclocal}
\ee
where we have taken $R_{\rm WD} = 10^{9}M_{\odot}$.  

Figure \ref{fig:burning} in Appendix \ref{sec:1D} compares $t_{\rm visc}$ and $t_{\rm nuc}$ for a few key nuclear reactions as a function of radius using a 1D steady-state disk model.  Near the outer edge of the disk $\sim R_{\rm d,0}$ (Eq.~\ref{eq:Rc}) where $T_0 \lesssim 10^{7}$ K (Eq.~\ref{eq:T0}), the timescale for hydrogen burning exceeds the age of the universe.  Even close to the WD surface $r \sim R_{\rm WD}$, where the temperature can reach $\gtrsim 10^{8}$ K for massive WDs, the hydrogen burning timescale is orders of magnitude larger than the inflow time.  A similar conclusion holds for the burning channel responsible for lithium production, $^{3}$He + $^{4}$He $\rightarrow ^{7}$Be + $\gamma$.  As a result of these findings (and confirmed by our full hydrodynamical simulations), we conclude that the disk outflows will be largely unprocessed, i.e. of approximately solar metallicity composition if the companion was on the main sequence prior to disruption.

As we discuss in Section \ref{sec:shell}, significant nucleosynthesis will occur over much longer timescales $\gg t_{\rm visc,0}$ in the hydrostatic hydrogen burning shell that accumulates on the WD surface.  Outflow from this envelope could in principle ``pollute'' the earlier disk wind ejecta with shell burning products, imparting the gaseous merger nebula with a non-solar metallicity signature (Section \ref{sec:CKVul}).

\section{Hydrodynamical Simulations}
\label{sec:numerical}

\begin{table*}
  \begin{center}
    \caption{Simulation Suite}
    \label{tab:table1}
    \begin{tabular}{c|c|c|c|c|c|c|c} 
      Model & $M_{\rm WD}$ & $R_{\rm WD}$ & $M_{\star}$ & $R_{\star}$ & $\alpha$ & $R_{\rm d,0}$ & $t_{\rm visc,0}^{(a)}$ \\
      \hline 
- & ($M_{\odot}$) & (cm) & ($M_{\odot}$) & ($R_{\odot}$) & - & (cm) & (s) \\
\hline
\hline 
\texttt{A0} & 0.3 & $1.2\times 10^{9}$ & 0.10 & 0.17 & 0.1 & $2.9\times 10^{10}$ & $7.3\times 10^{4}$ \\
\texttt{A1} & 0.6 & 10$^{9}$ & 0.20 & 0.28 & 0.1 & $5.0\times 10^{10}$ & $1.1\times 10^{5}$ \\
\texttt{A2} & - & - & - & - & 0.01 & - & $1.1\times 10^{6}$\\
\texttt{A3} & 0.66 & $8.39\times 10^{8}$ & 0.25 & 0.34 & 0.1 & $5.6\times 10^{10}$ & $1.2\times 10^{5}$\\
\texttt{A4} & 0.52 & $9.41\times 10^{8}$ & 0.20 & 0.28 & - & $4.6\times 10^{10}$ & $1.1\times 10^{5}$\\
\texttt{A5} & 1.09 & $4.66\times 10^{8}$ & - & - & - & $6.6\times 10^{10}$ & $1.3\times 10^{5}$ \\
\hline
    \end{tabular}
\label{tab:models}\\
$^{(a)}$Calculated from Eq.~(\ref{eq:tvisc}), assuming a disk aspect ratio $\theta = 0.33$.
  \end{center}
\end{table*}

\begin{table*}
  \begin{center}
    \caption{Simulation Results$^{\dagger}$}
    \label{tab:table1}
    \begin{tabular}{c|c|c|c|c|c} 
      Model & $M_{\rm ej}(E>0$) & $M_{\rm acc}(E < 0; \Omega < 0.2 \Omega_{\rm K})$ & $\langle v_{\rm ej} \rangle$ & $\langle \mathcal{S}/v_{\rm ej}^{2} \rangle$ & $ M_{\rm ej}(\mathcal{B}e>0)^{\ddagger}$ \\
      \hline 
- & ($M_{\odot}$) & ($M_{\odot}$) & (km s$^{-1}$) & (K s$^{2}$ g$^{-2/3}$) & ($M_{\odot}$) \\
\hline
\hline 
\texttt{A0} & 0.072 &$1.0\times 10^{-2}$ & 791 & $ 2.1\times 10^{-6}$ & 0.089\\
\texttt{A1} & 0.183 & $7.7\times 10^{-3}$ & 648 & $ 3.3\times 10^{-6}$ & 0.191\\
\texttt{A2} & 0.181 & $7.2\times 10^{-3}$ & 612 & $ 1.8\times 10^{-6}$ & 0.190\\
\texttt{A3} & 0.207 & $2.4\times 10^{-2}$ & 764 & $1.2\times 10^{-6}$& 0.220\\
\texttt{A4} & 0.185 & $6.8\times 10^{-3}$ & 805 & $2.3\times 10^{-6}$& 0.191\\
\texttt{A5} & 0.180 & $1.1\times 10^{-2}$ & 725 & $1.5\times 10^{-6}$& 0.186\\

\hline
    \end{tabular}
\label{tab:results}
  \end{center}
  
    $^{\dagger}$All results quoted at $t = 1 t_{\rm visc,0}$ to allow a direct comparison.  Average quantities $\langle... \rangle$ are associated with the unbound ejecta and are weighted by mass. $\ddagger$ Alternative definition of ejecta mass as material with positive Bernoulli parameter $\mathcal{B}e = E + P/\rho$.
    \end{table*}

This section describes hydrodynamical simulations of the post-merger disk evolution.  Our main goal is to quantify the properties of the disk outflows (quantity, velocity, angular distribution) and of the matter which ends up in a bound spherical envelope on the WD, as these inform the short- and long-term electromagnetic signatures of the merger, respectively (Section \ref{sec:transient}).  

We do not simulate the dynamical disruption phase itself.  Instead, following previous works (e.g., \citealt{Metzger12,Schwab+12,Fernandez+2013,Zenati19a,Zenati+20}), we start our simulations after the disk has formed and focus on the system evolution that occurs on longer, viscous timescales, over which the geometry will be approximately azimuthally symmetric.   

\subsection{Simulation Setup}

Our simulations are performed using FLASH, an adaptive mesh refinement (AMR) code that solves the hydrodynamic Euler equations using an unsplit piecewise-parabolic method (${\rm PPM}$) solver \citep{Fryxell2000}.  We use ${\rm 2D}$ axisymmetric cylindrical coordinates $[{\bar\rho},\phi,z]$ on a grid of size $(0.2-8\times 10^{13}$ cm$)\times (0.7-6\times 10^{13}$ cm), significantly larger than the orbital size of the CV binary prior to the merger.
 
We solve the equations of conservation of mass, momentum, energy, and chemical species,
\begin{eqnarray}
\frac{\partial\rho}{\partial t}+\nabla\cdot\left(\rho\mathbf{v}\right)&=&0,  \\
\frac{d\mathbf v}{dt}&=&\mathbf{f}_{c}-\frac{1}{\rho}\nabla p + \nabla\phi,  \\
\rho \frac{dl_z}{dt} &=& \bar\rho (\nabla\cdot\mathbf{T})_\phi \\
 \rho\frac{d e_{\rm int}}{d t}+p\nabla\cdot\mathbf v &=& \frac{1}{\rho \nu}\mathbf{T}:\mathbf{T}+\rho(\dot{Q}_{\rm nuc} - \dot{Q}_\nu), \\
 \frac{\partial \mathbf{X}}{\partial t} &=& \dot{\mathbf{X}}, \label{eq:chemical_species_evolution} \\
 \nabla^{2}\phi&=& 4 \pi G \rho + \nabla^{2}\phi_{\rm c}, \\
\mathbf{f}_{c}&=&\frac{l_{z}^{2}}{\bar\rho^{3}}\hat{{\bar\rho}},
\end{eqnarray}
where $d/dt \equiv \partial/\partial t + \mathbf{v}\cdot\nabla$.  Variables have their standard meaning: $\rho$, $\mathbf{v}$, $p$, $e_{\rm int}$, $\nu$, $\mathbf{T}$, $\phi$, and $\mathbf{X}=\{X_i\}$ denote, respectively, fluid density, poloidal velocity, total pressure, specific internal energy, fluid viscosity, viscous stress tensor for azimuthal shear, gravitational potential, and mass fractions of the isotopes $X_i$, with $\sum_i X_i = 1$.  
The quantity $\mathbf{f}_{c}$ is an implicit centrifugal source term, where $l_z$ is the z-component of the specific angular momentum.  The gravity of the central WD is represented as a point-mass gravitational potential to the solver, $\phi_{\rm c}$.  Self-gravity is included as a multipole expansion of up to $l_{\max}=48-60$ terms using the new FLASH multipole solver and the super time-steps (STS) method for calculating the adaptive time-steps (\texttt{FLASH4.6.2}).  

The quantities $\dot{Q}_{\rm nuc}$ and $\dot{Q}_{\rm \nu}$ represent the specific heating rate due to nuclear reactions and the specific cooling rate due to neutrino emission, respectively.  We do not include nuclear burning in our presented simulations.  We have checked that this is a good assumption by comparing our simulation results without burning to a test case with nuclear burning activated, finding no significant differences in the evolution (see also Section \ref{sec:burning} and Fig.~\ref{fig:burning}).  We include neutrino cooling in the internal energy evolution \citep{Houck1991}, although it also has no appreciable impact given the relatively low temperatures reached in the accretion flow.

We employ the Helmholtz equation of state \citep{Timmes2000}, which includes gas pressure, radiation pressure, and degeneracy pressure.  The equation of state does not include hydrogen recombination energy, which is not dynamically important in the disk or outflow launching region (though it does play a crucial role in the post-merger optical transient; Section \ref{sec:transient}).

Our simulations do not include magnetic fields, so we cannot self-consistently account for angular momentum transport due to the MRI.  Due to the axisymmetric nature of the simulations, we also cannot capture the dominant $m =1$ instability arising due to self-gravity.  We model both of these processes in an approximate way, by employing a kinematic $\alpha$-viscosity of the standard form (see also Eq.~\ref{eq:tvisc}),
\begin{equation} 
    \nu_{\alpha}=\alpha c_{s}^{2}/\Omega_{\rm K},  \label{eq:nualpha}
\end{equation}
where $\Omega_{\rm K} = (GM_{\rm WD}/r^{3})^{1/2}$ is the Keplerian frequency and $c_{s}$ the sound speed.  We explore the sensitivity of our results to different values of $\alpha = 0.01,0.1$ (Table \ref{tab:table1}).

We treat the inner WD surface as a hard surface by employing a reflecting condition at the inner boundary $R_{\rm in} = R_{\rm WD}$.  We apply outflow conditions to the outer boundary of the simulation domain.   We follow the system evolution for several viscous timescales at the initial outer disk radius $R_{\rm d,0}$, until most of the original star has either been accreted onto the WD or unbound from the system in outflows.

The initial conditions of the torus are set up following the procedure described in \citet{Fernandez+2013} and \cite{Zenati19a, Zenati19b}, with the total initial torus mass equal to that of the star, and its total energy equal to that of the binary prior to disruption.  We self-consistently relax the initial torus before turning on the viscosity following the iterative method described in \citet{Zenati19a}.  The viscous spreading from the $\alpha$-viscosity term subsequently produces a disk whose properties quickly resemble those generated from the disruption of the star.  We employ a spatial resolution of $6-20$ km, which we found is sufficient to achieve $\lesssim$ 10$\%$ conservation in energy.

\subsection{Simulation Results}

The suite of simulations and their key properties are summarized in Table \ref{tab:models} and Table \ref{tab:results}, respectively.  The latter includes the total ejecta mass ($M_{\rm ej}$, defined as matter with positive specific energy $ E > 0$), total mass accreted by the WD ($M_{\rm acc}$, which we define as gravitationally bound matter with $\Omega/\Omega_{\rm K} < 0.2$, i.e. modest centrifugal support), mass-averaged velocity of the unbound ejecta ($\langle v_{\rm ej}\rangle$),  and a mass-averaged ``entropy-like'' quantity of the unbound ejecta ($\mathcal{S}/v_{\rm ej}^{2}$, where $\mathcal{S} \equiv T/\rho^{2/3}$; Eq.~\ref{eq:S}).  The quantity $\mathcal{S}/v_{\rm ej}^{2}$, which is conserved in the outflow once it stops being heated by viscosity and accelerating out of the WD gravitational potential well, defines an initial condition for calculating the optical transient emission from the ejecta (Appendix \ref{sec:lightcurve}).  

The final column of Table \ref{tab:results} also provides a second definition of the mass of unbound ejecta, as matter with positive Bernoulli parameter ($\mathcal{B}e \equiv E + P/\rho$).  Although matter with $E < 0$ but $\mathcal{B}e > 0$ is not unbound by the final snapshot of the simulation, its enthalpy is in principle high enough that it could become unbound by pressure forces at a later point.  This distinction makes little practical difference: ejecta masses defined by $E > 0$ agree with those defined by $\mathcal{B}e > 0$ in all models to $< 10\%$.

\begin{figure*}
\includegraphics[width=\linewidth]{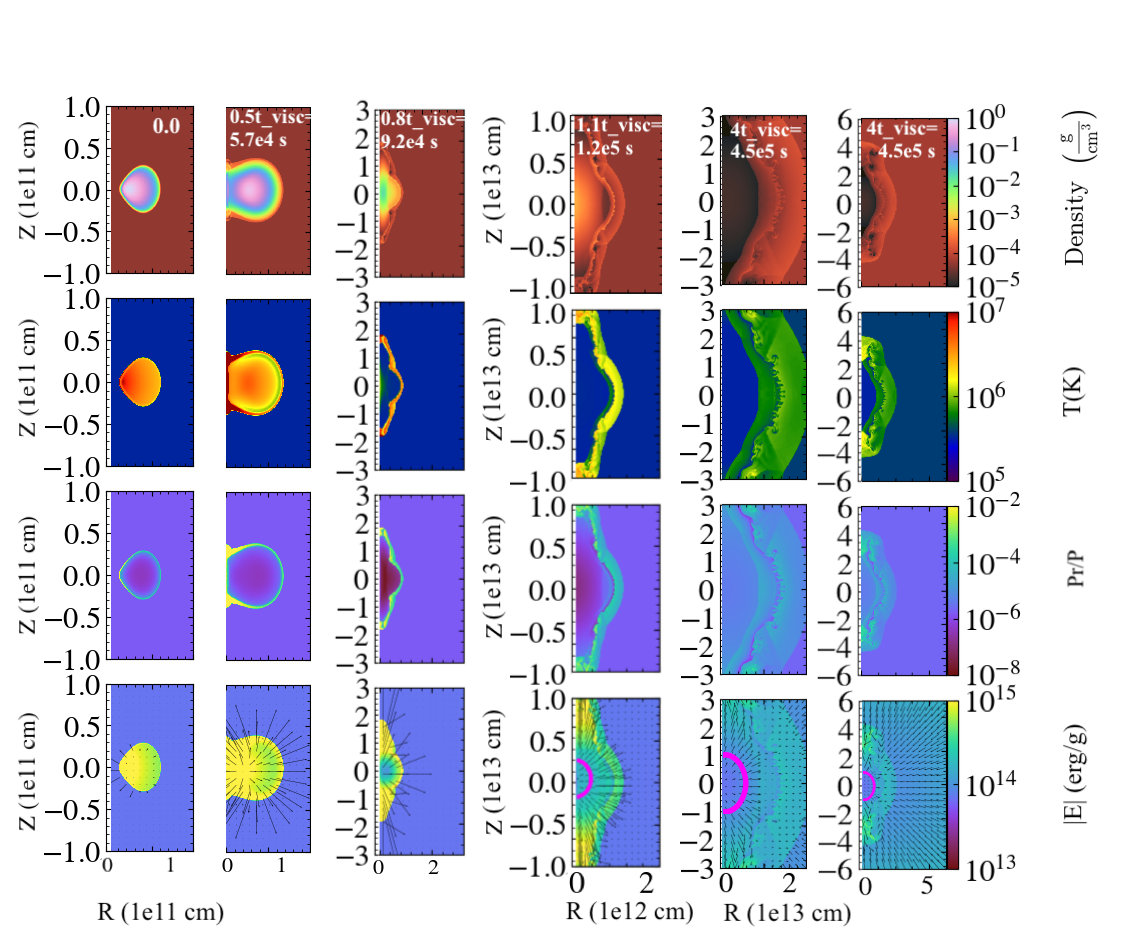}
\caption{Snapshots of model \texttt{A1} ($M_{\rm WD}=0.6M_{\odot}$; $M_{\star} = 0.2\,$M$_\odot$; $\alpha = 0.1$) showing the accretion of the post CV-merger torus onto the central WD (located at the origin $R = Z = 0$).  The timescale of each snapshot is marked along the top of each column in seconds as well as the initial viscous time at the outer edge of the torus, $t_{\rm visc,0}$ (Eq.~\ref{eq:tvisc}).  From top to bottom, rows show: density, $\rho$; temperature, $T$; ratio of radiation pressure to total pressure, $P_{\rm rad}/P$, and the absolute value of the total (kinetic + gravitational + thermal) specific energy, $|E|$.  The pink contour in the bottom row separates the unbound ejecta ($E>0$) from the spherical accreted envelope that remains gravitationally bound to the WD ($E < 0$).  Note the difference in radial scale between the columns, with the final two snapshots showing the same epoch but at different spatial scales.}
\label{fig:snapshots}
\end{figure*}

We focus on describing results for the fiducial model \texttt{A1} ($M_{\rm WD} = 0.6M_{\odot}$; $M_{\star} = 0.2M_{\odot}$; $\alpha = 0.1$), as the other models exhibit similar qualitative evolution.  Figure \ref{fig:snapshots} shows properties of the disk as viewed through through the midplane at several snapshots in time ranging from the initial state to $t = 4t_{\rm visc,0} \approx 5\times 10^{5}$ s $\sim 1$ week.  The inner edge of the torus begins accreting onto the WD within a small fraction $\lesssim 0.1$ of the viscous timescale defined at $R_{d,0} \sim R_{\odot}$, $t_{\rm visc,0}$ (Eq.~\ref{eq:tvisc}).  Gravitational energy released by the inflow, unable to cool through radiation, generates regions of high pressure that lead to the ejection from the inner disk of material along the polar axis at velocities $\sim 1000$ km s$^{-1}$ (see Eq.~\ref{eq:vfast}).  

Over longer timescales, the disk begins to viscously spread outwards in radius as a result of the redistribution of angular momentum by viscosity (see below) and thicken due to viscous heating.  Over this phase, matter is also unbound in outflows directed closer to the equatorial plane.  The outflow velocities are shown as arrows in the final row of Fig.~\ref{fig:snapshots}.  The pink contour in the final snapshots defines the $E = 0$ surface which separates the aspherical unbound ejecta shell ($E > 0$) from the quasi-spherical envelope that remains gravitationally bound to the WD ($E < 0$).  Both the disk and the outflows remain gas-pressure supported at all radii and times (Fig.~\ref{fig:snapshots}), as expected from analytic estimates (Eq.~\ref{eq:Pratio}) and 1D time-independent model (Fig.~\ref{fig:burning}, top panel).

\begin{figure*}
\includegraphics[width=\linewidth]{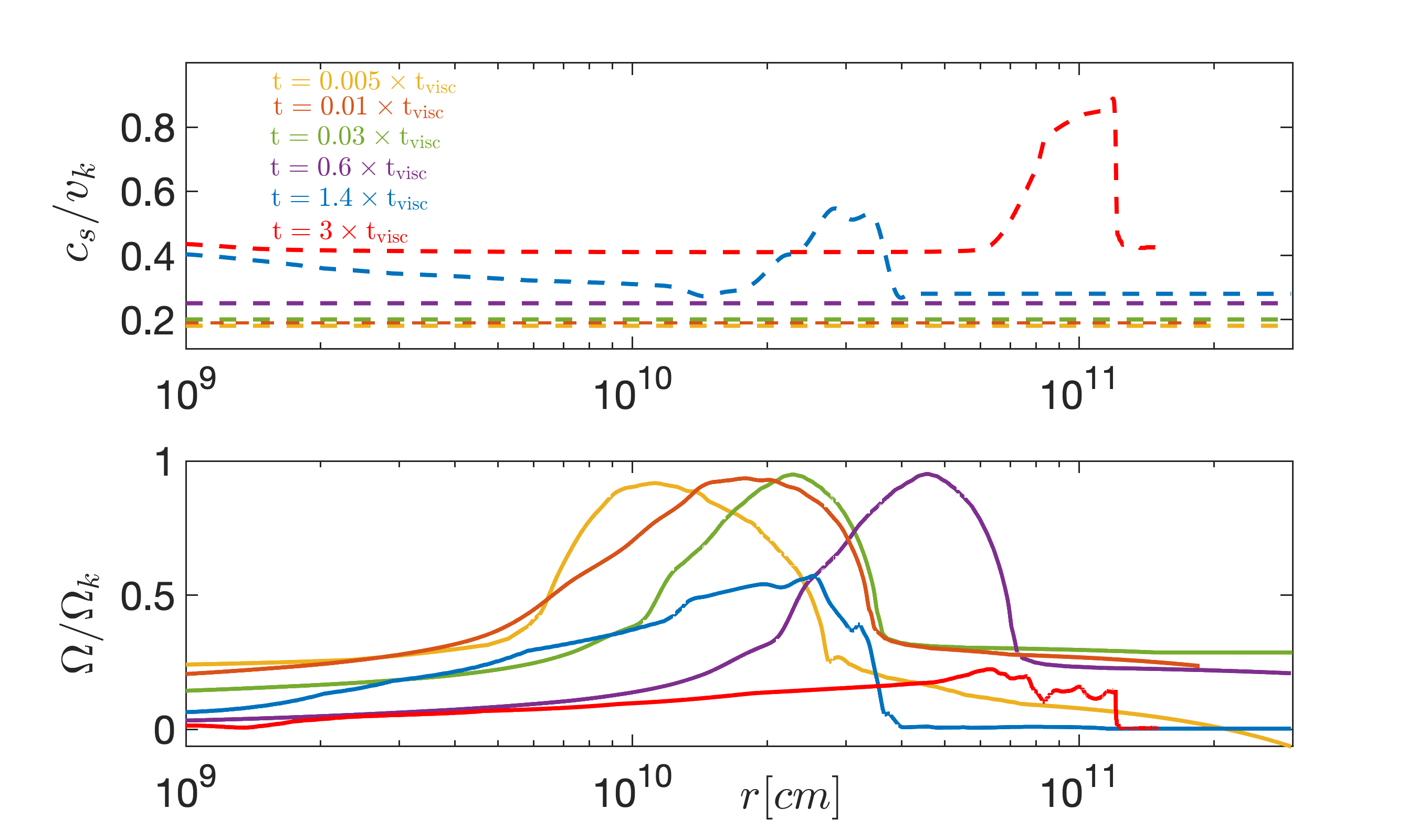}
\caption{Evolution of the merger debris from a rotationally supported disk immediately following the disruption to a pressure supported envelope on the WD surface over several viscous times.  Here we are showing several snapshots from our fiducial model \texttt{A1} of the (angle-averaged) radial profile of $\Omega/\Omega_{\rm K}$ (lower panel) and $c_{\rm s}/v_{\rm K}$ (upper panel).  These quantities show the importance of rotational and pressure support, respectively, where $\Omega_{\rm K}$ is the Keplerian angular rotation rate, $v_{\rm K} = r \Omega_{\rm K}$, and $c_{\rm s}$ the sound speed.  }
\label{fig:rotational_support}
\end{figure*}

In support of the above picture, Figure \ref{fig:rotational_support} shows snapshots of the angle-averaged radial profiles of the angular velocity $\Omega/\Omega_{\rm K}$ and virial parameter $c_{\rm s}/v_{\rm K}$.  We observe the transformation, over several viscous times, from a rotation-supported torus at $r \sim few \times 10^{10}$ cm $\lesssim R_{\rm d,0}$ ($\Omega/\Omega_{\rm K} \sim 1$; $c_{\rm s}/v_{\rm K} \ll 1$) to a pressure-supported envelope ($c_{\rm s} \sim 0.5 v_{\rm K}; \Omega/\Omega_{\rm K} \ll 1$) surrounding the WD, with the remnant torus now spreading to larger radii ($r \sim 10^{11}$ cm) through a combination of viscosity and radial pressure forces.

Figure \ref{fig:allmass} shows the cumulative mass with time accreted by the WD and unbound in outflows, the latter defined separately according to the positive energy or positive Bernoulli parameter criterion.  By either definition, we observe that within less than a viscous time at the outer disk radius $t_{\rm visc,0} \sim$ 1 day (or several viscous times defined at smaller radii), most of the original torus is ejected ($M_{\rm ej} \simeq 0.19M_{\odot}$, while a significant mass $\gtrsim 7\times 10^{-3}M_{\odot}$ has accumulated into a quasi-spherical envelope around the WD. 

\begin{figure*}
\includegraphics[width=\linewidth]{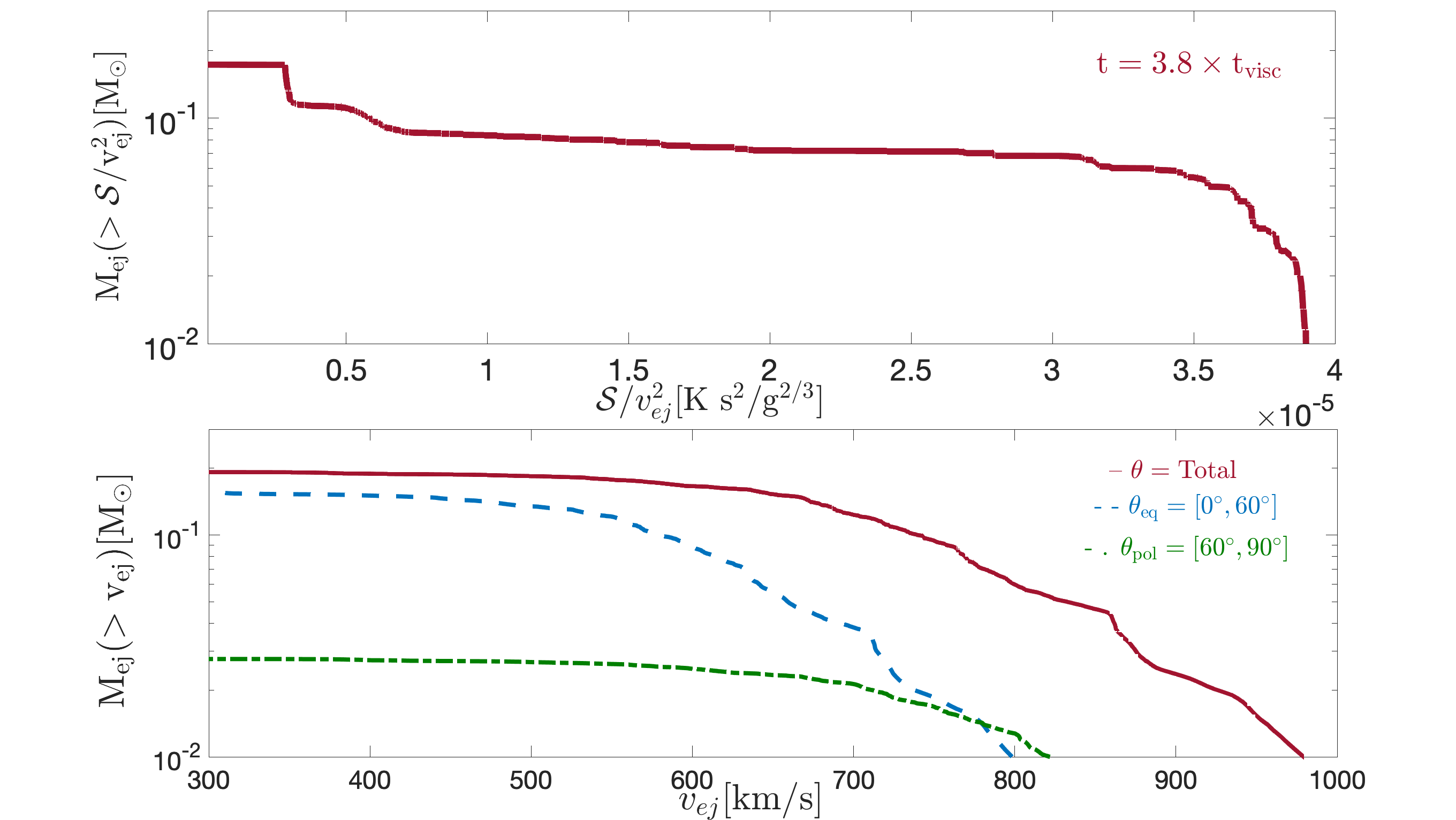}
\caption{{\it Top Panel:} Cumulative mass distribution for the unbound material, plotting the amount of mass above a given value of the internal energy ($\mathcal{S}/v_{\rm ej}^{2}$, where $\mathcal{S} \equiv T/\rho^{2/3}$ [Eq.~\ref{eq:S}]). {\it Bottom Panel:} Cumulative mass distribution of the unbound ejecta above a given velocity ($v_{\rm ej}$). Both panels are calculated for model \texttt{A1} at a late snapshot $t = 4 t_{\rm visc,0}$ when these distributions are effectively frozen.  In the bottom panel, the velocity distribution is further divided into ejecta focused into the equatorial plane of the binary (blue dashed line; polar angle $\theta \in [0,60^{\circ}]$ measured from the equator) and along the polar axis (green dot-dashed line; $\theta \in [60^{\circ},90^{\circ}]$).}
\label{fig:vel_distribution}
\end{figure*}

\begin{figure}
\includegraphics[width=\linewidth]{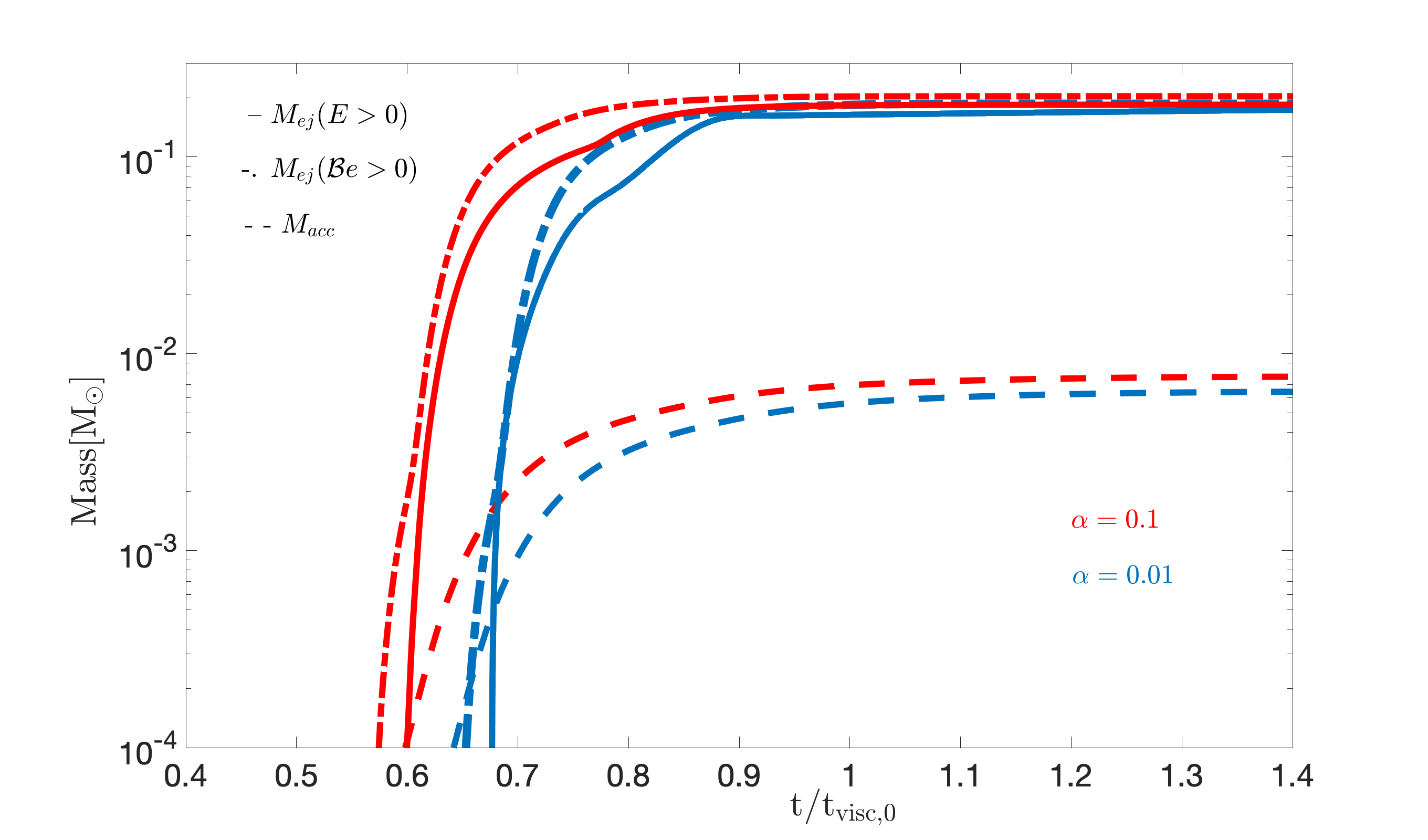}
\caption{Evolution of the accreted mass and unbound ejecta mass as a function of time in units of the initial viscous timescale (Eq.~\ref{eq:tvisc}) for model \texttt{A1} (red curves) and \texttt{A2} (blue curves).  We show the unbound ejecta calculated two ways, according to positive total energy ($E>0$; solid line) and positive Bernoulli parameter ($\mathcal{B}e>0$; dot-dashed line).  The bound mass $(E < 0)$ accreted into a pressure-supported envelope on the WD surface ($\Omega < 0.2\Omega_{\rm K}$) are shown as dashed lines. }
\label{fig:allmass}
\end{figure}

The bottom panel of Figure \ref{fig:vel_distribution} shows the mass of the ejecta $M_{\rm ej}(> v_{\rm ej}$) above a given velocity $v_{\rm ej}$, broken down separately into the matter ejected into the polar direction (velocity vector within 30$^\circ$ of the rotational axis) and that released into the equatorial direction.  About 15\% of the ejecta emerges in the polar region, with velocities up to $\sim 1000$ km s$^{-1}$.  However, most of the mass loss occurs at lower latitudes, with a lower average velocity of $\sim 600-700$ km s$^{-1}$, in agreement with analytic estimates (Eq.~\ref{eq:vw}).  

Although our discussion has been focused on the fiducial \texttt{A1} model, we do not see a significant difference in the quantitative results if we decrease the viscosity from $\alpha = 0.1$ to $\alpha = 0.01$ (model \texttt{A2}), other than an overall slower evolution of the system by a factor $\sim 10$ (Table \ref{tab:results}).  We also find little appreciable qualitative changes for different WD masses (models \texttt{A0}, \texttt{A3}-\texttt{A5}).  In model \texttt{A0} of the merger of 0.1$M_{\odot}$ with a $0.3M_{\odot}$ WD, we find ejecta and accreted masses of $\approx 0.09M_{\odot}$ and $\approx 0.01M_{\odot}$, respectively, and a mean ejecta velocity $\langle v_{\rm ej}\rangle \approx 790$ km s$^{-1}$.

\section{Transient Emission}
\label{sec:transient}

Several sources of optical/IR transient emission will follow the CV merger, on timescales ranging from days to millions of years.   

\subsection{Recombination-Powered Optical/IR Transient}

As discussed in Sections \ref{sec:analytic} and \ref{sec:numerical}, the accretion phase after the merger results in the ejection of a significant mass $M_{\rm ej} \sim M_{\star} \sim 0.1-1M_{\odot}$ at a characteristic velocity $v_{\rm ej} \approx \langle v_{\rm w} \rangle$ of several hundred km s$^{-1}$ (Eq.~\ref{eq:vw}; Fig.~\ref{fig:vel_distribution}).  No significant nuclear burning takes place in the disk (Section \ref{sec:burning}; Fig.~\ref{fig:burning}), so the composition of the ejecta will be approximately solar.  The ejecta mass and average velocity broadly overlap those inferred from LRN transients associated with binary star mergers (e.g., \citealt{Blagorodnova+21}), though with a greater fraction of the ejecta potentially attaining higher velocities $\gtrsim 10^{3}$ km s$^{-1}$ due to the deeper gravitational potential of the WD.

Immediately after leaving the disk, the ejecta is highly optically thick, with its radiation trapped in the flow.  However, as the expanding material dilutes, the optical depth through it decreases, eventually enabling optical wavelength emission to escape.  This occurs on the photon diffusion timescale, which also defines the rise time of the light curve and can be approximately written as (e.g., \citealt{Arnett82})
\begin{eqnarray}
t_{\rm pk} &\approx& \left(\frac{M_{\rm ej}\kappa}{4\pi v_{\rm ej}c}\right)^{1/2} \nonumber \\
&\approx& 66\,{\rm d}\,\left(\frac{M_{\rm ej}}{0.3\,M_{\odot}}\right)^{1/2}\left(\frac{v_{\rm ej}}{500\,{\rm km\,s^{-1}}}\right)^{-1/2}\left(\frac{\kappa}{1\,{\rm cm^{2}\,g^{-1}}}\right)^{1/2} \nonumber \\
\label{eq:tpk}
\end{eqnarray}
where $\kappa$ is the opacity, normalized to a typical value during the epoch when the ejecta is becoming diffusive (e.g., \citealt{Metzger&Pejcha17}).  Notice that $t_{\rm pk}$ (Eq.~\ref{eq:tpk}) is longer than the disk outflow time timescale $\sim t_{\rm visc,0}$ (Eq.~\ref{eq:tvisc}).  This justifies treating the wind ejecta, on timescales $t \gtrsim t_{\rm pk}$, as residing within a single expanding shell of radial thickness $\sim v_{\rm ej}t$, instead of a continuous wind (this assumption is implicit in Eq.~\ref{eq:tpk}).  The shell thickness is set by the internal velocity dispersion, $\delta v_{\rm ej} \sim v_{\rm ej}$ (Fig.~\ref{fig:vel_distribution}).

\begin{figure}
\includegraphics[width=\linewidth]{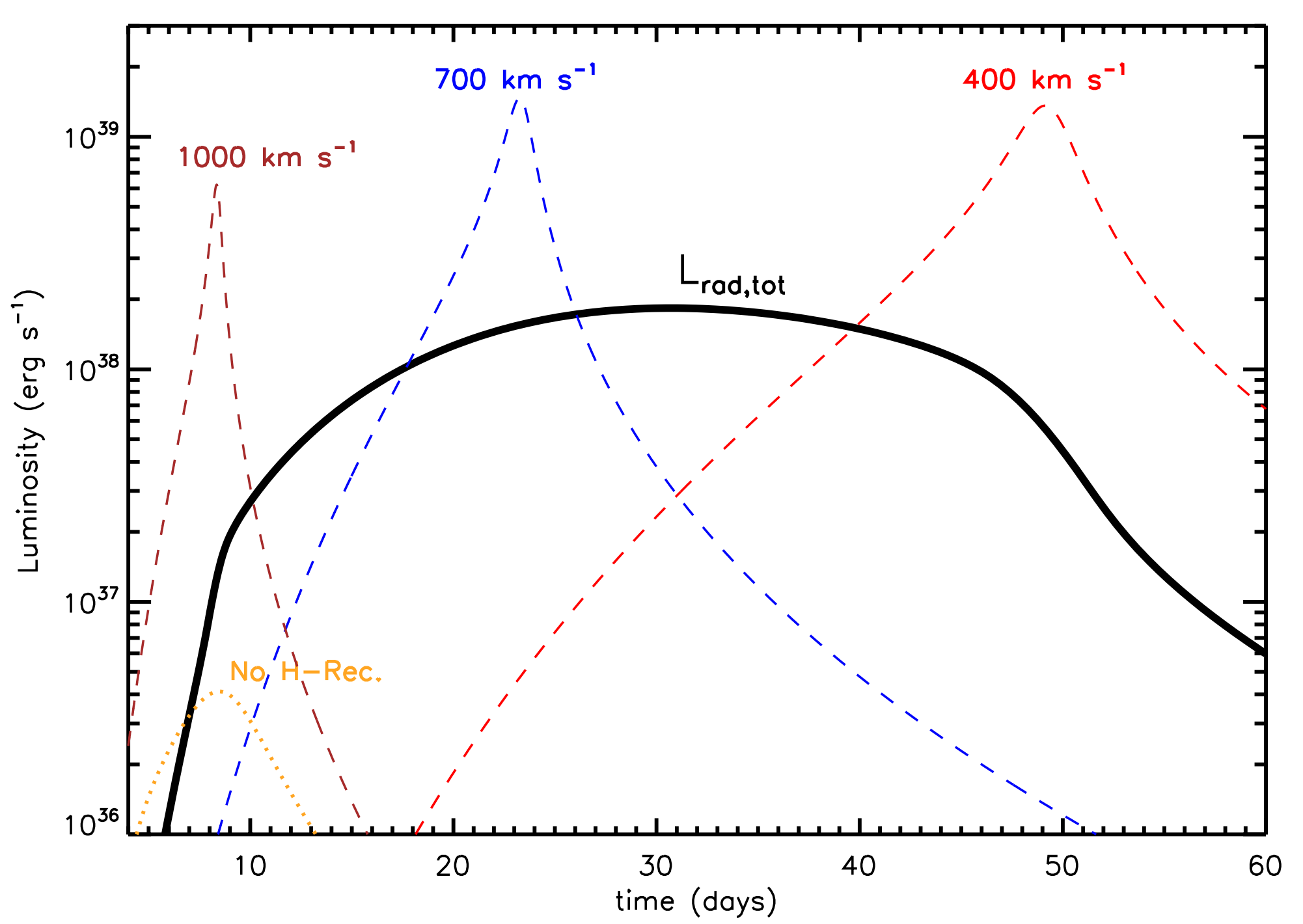}
\caption{Optical light curve of the CV merger transient, calculated following the semi-analytic model described in Appendix \ref{sec:lightcurve} and assuming ejecta parameters $M_{\rm ej} = 0.4M_{\odot}$, $R_{\rm d,0} = 1 R_{\odot}$, and $\eta = 100$ motivated by our hydrodynamical simulations (here $\eta \equiv E_0/(M_{\rm ej}v_{\rm ej}^{2})$ is a dimensionless internal energy, $E_0$, of the ejecta; Eq.~\ref{eq:eta}).  Dashed lines show the result of a one-zone (single ejecta velocity) model for different values $v_{\rm ej} = 400, 700, 1000$ km s$^{-1}$ as marked.  A solid line shows the more physical multi-zone model characterized by a distribution of ejecta velocities, as motivated by the results of our numerical simulation (Eq.~\ref{eq:Menc}; Fig.~\ref{fig:vel_distribution}).  An orange line shows the light curve obtained if we assume adiabatic evolution, i.e.~ignoring the energy released by hydrogen recombination.}
\label{fig:lightcurve}
\end{figure}

Due to the compact nature of the disk ($\lesssim R_{\odot}$; e.g., relative to the red giant progenitors of Type IIP supernovae), the thermal energy of the wind ejecta will experience large adiabatic losses before expanding to the point that light can escape and carry this energy to a distant observer.  As a result, the dominant source powering the transient's luminosity is not the initial heat carried out from the disk by the outflows, but instead the energy released at larger radii by hydrogen recombination (we show this below in Fig.~\ref{fig:lightcurve} by comparing a light curve calculated with and without recombination effects).  The recombination energy is given by $E_{\rm rec} \simeq (XM_{\rm ej}/m_p)E_{\rm Ryd} \approx 2\times 10^{45}(M_{\rm ej}/0.1M_{\odot})$ erg, where $X \approx 0.74$ is the hydrogen mass fraction and $E_{\rm Ryd} = 13.6$ eV the Rydberg energy. 

The peak luminosity of the optical transient emission can therefore be roughly estimated as
\begin{eqnarray}
L_{\rm pk} &\sim& \frac{f_{\rm ad} E_{\rm rec}}{t_{\rm pk}} \approx 3\times 10^{38}{\rm erg\,s^{-1}}\left(\frac{f_{\rm ad}}{0.3}\right) \times \nonumber \\
&& \left(\frac{M_{\rm ej}}{0.3M_{\odot}}\right)^{1/2}\left(\frac{v_{\rm ej}}{500\,{\rm km\,s^{-1}}}\right)^{1/2}\left(\frac{\kappa}{1\,{\rm cm^{2}\,g^{-1}}}\right)^{-1/2},
\label{eq:Lpk}
\end{eqnarray}
where the factor $f_{\rm ad} < 1$ accounts for the partial loss of the thermal energy released due to $PdV$ work prior to the opacity dropping enough to allow radiation to escape.  Recombination occurs when the ejecta temperature $T \approx 10^{4}$ K and the opacity is still very large, i.e. at times prior to $t_{\rm pk}$.  We have normalized $f_{\rm ad} \sim 0.3$, based on the approximate value found in our light curve model (see below).

In Appendix \ref{sec:lightcurve} we present a model for the recombination-powered light curve, which accounts for the various relevant sources of opacity in the ejecta and takes as input the distributions of velocity and internal specific energy of the unbound ejecta as determined from our numerical simulations.  The black curve in Figure \ref{fig:lightcurve} shows the resulting light curve for fiducial assumptions ($M_{\rm ej} = 0.4 M_{\odot}$, $v_{\rm ej} \approx 400-1000$ km s$^{-1}$).  The light curve exhibits a plateau shape lasting about a month at a luminosity of a few $10^{38}$ erg s$^{-1}$, with the fastest expanding layers contributing to the earliest emission (and the slowest layers to the latest emission).  

In summary, the predicted luminosities $\gtrsim 10^{38}-10^{39}$ erg s$^{-1}$ and timescales $\sim $ months of CV merger transients are comparable to those of slow classical novae and LRN from stellar mergers (e.g., \citealt{Bond+03,Pastorello+19,Blagorodnova+21}).

\subsection{Circumstellar Interaction and Dust Formation}
\label{sec:dust}

Observations of LRN from stellar mergers prior to optical peak (e.g., \citealt{Tylenda+11}) indicate that substantial mass-loss occurs from the binary in the final phases of runaway mass-transfer leading to the merger (\citealt{Pejcha+17}), likely in the form of outflows from the $L_{2}$ Lagrange point (e.g., \citealt{Pejcha+16b,Pejcha+16a}).  Similar pre-dynamical mass loss may occur leading up to a CV merger.  Furthermore, if the CV is driven to merge as a result of frictional drag from classical novae (e.g., \citealt{Shen15,Schreiber+16}), then the nova shell or circumbinary disk (of mass $\sim 10^{-5}-10^{-4}M_{\odot}$) could still be engulfing the binary at the time of merger.  

If such a dense gaseous medium extends to large radii around the binary at the time of the merger, this could give rise to a second source of emission powered by the shock interaction with the merger ejecta (e.g., \citealt{Metzger&Pejcha17} for a detailed model).  This will enhance the optical luminosity compared to the estimate in Eq.~(\ref{eq:Lpk}) and lengthen its duration if the external medium extends to large radii $\gg t_{\rm p}v_{\rm ej} \sim 1-10$ AU.  Shock interaction could also manifest spectroscopically as emission lines from hot post-shock gas or from the irradiated upstream medium, as observed in Type IIn supernovae (e.g., \citealt{Smith+08}). 

The expanding ejecta will cool to temperatures $\lesssim 1000-2000$ K, starting in the outer layers on a timescale $\sim t_{\rm pk}$ (Eq.~\ref{eq:tpk}),  enabling the formation of molecules and dust, as observed in LRN (e.g., \citealt{Kaminski+10}), classical novae (e.g., \citealt{Gehrz+98}), and CK Vul (e.g., \citealt{Eyres+18,Kaminski+20a}; Section \ref{sec:CKVul}).  After dust forms and blocks optical light from the central source, the spectral energy distribution of the emission will shift into the infrared bands.  The dust will eventually become optically thin, but only after a long timescale
\begin{eqnarray}
&&t_{\rm thin} \sim \left(\frac{\kappa_{\rm d}M_{\rm ej}}{4\pi v_{\rm ej}^{2}}\right)^{1/2} \nonumber \\
&\approx& 43\,{\rm yr}\left(\frac{M_{\rm ej}}{0.3M_{\odot}}\right)^{1/2}\left(\frac{\kappa_{\rm d}}{10^{2}\,\rm cm^{2}\,g^{-1}}\right)^{1/2}\left(\frac{v_{\rm ej}}{500\,{\rm km\,s^{-1}}}\right)^{-1/2}, \nonumber \\
\label{eq:tthin}
\end{eqnarray}
where $\kappa_{\rm d}$ is the opacity of the dust at optical wavelengths.  Thus, it may take decades for the central remnant (described in the next section) to become visible at optical wavelengths.

\subsection{Central Remnant}
\label{sec:shell}

The portion of the secondary not unbound in disk outflows will end up as a spherical shell on the WD surface, with an estimated mass $M_{\rm acc} \sim 0.02M_{\star} \sim 10^{-3}-10^{-2}M_{\odot}$ (Eq.~\ref{eq:Macc}; Fig.~\ref{fig:allmass}).  The base of this shell will become hot enough to undergo hydrogen burning.  As long as the layer mass exceeds a critical minimum value ($M_{\rm acc} \sim 10^{-4}M_{\odot}$ for a WD of mass $M_{\rm WD} \lesssim 0.5-0.6M_{\odot}$; e.g., \citealt{Shen&Bildsten07,Nomoto+07}), it will expand to giant dimensions $R \sim 1$ AU on the thermal timescale $t_{\rm th} \sim GM_{\rm WD}^{2}/(R L_{\rm shell}) \sim 10$ yr.  Here, $L_{\rm shell}$ is the steady-state luminosity of the hydrogen burning shell (\citealt{Paczynski70}),
\begin{eqnarray}
L_{\rm shell} \approx 2\times 10^{38}\,{\rm erg\,s^{-1}}\,\left(\frac{M_{\rm WD}}{M_{\odot}}-0.522\right), \,\,\, \nonumber \\ 0.57 \lesssim M_{\rm WD}/M_{\odot} \lesssim 1.39 
\label{eq:Lshell}
\end{eqnarray}
For WDs of lower masses $M_{\rm WD} \approx 0.4(0.5)M_{\odot}$, the shell luminosity is lower $L_{\rm shell} \approx 4\times 10^{36}(10^{37})L_{\odot}$. 

The shell luminosity $L_{\rm shell}$ for $M_{\rm WD} \approx 0.3-0.6M_{\odot}$ can be up to several orders of magnitude smaller than the transient generated during the transient mass ejection phase (Eq.~\ref{eq:Lpk}).  However, the timescale of shell burning,
\begin{eqnarray}
t_{\rm shell} &\sim& \frac{Q(XM_{\rm acc}/m_p)}{L_{\rm shell}} \nonumber \\&\approx&  3\times 10^{5}\,{\rm yr}\left(\frac{M_{\rm acc}}{10^{-2}M_{\odot}}\right)\left(\frac{L_{\rm shell}}{10^{37}{\rm erg\,s^{-1}}}\right)^{-1},
\label{eq:tshell}
\end{eqnarray}
is considerably longer, where $X \approx 0.74$ and $Q \simeq 6.4$ MeV per nucleon is the energy released by hydrogen burning.  Thus, even after the dusty ejecta shell becomes optically thin decades after the merger (Eq.~\ref{eq:tthin}), the central remnant will lie on the Hayashi track and remain red ($T_{\rm eff} \approx 3000$ K), with a significantly greater luminosity than its CV progenitor.

Equation (\ref{eq:tshell}) represents an upper limit on the duration of the shell burning phase insofar that it does not account for wind mass-loss from the remnant.  Such outflows will eventually pollute the promptly released ejecta shell, potentially endowing it with nuclear processed material.  Hydrogen burning in the shell will take place via the CNO cycle, and hence this region will be characterized by a nitrogen overabundance or other non-solar signatures (e.g., in the ratio $^{13}$C/$^{12}$C).  For massive WDs the temperature in the burning shell may become sufficiently high to synthesize radioactive $^{7}$Be (see dashed line in Fig.~\ref{fig:burning}), which could decay into $^{7}$Li after being transported outwards to cooler regions (e.g., \citealt{Cameron&Fowler71}).  If carried to the photosphere by mixing processes, resulting e.g. from an unstable entropy gradient imparted by the merger process, these burning products could provide distinguishing spectroscopic features of the stellar remnant (Section \ref{sec:rates}) or in the dusty/molecular nebula on larger scales (Section \ref{sec:CKVul}).   

\section{Discussion}
\label{sec:discussion}

\subsection{Rates and Remnant Populations}
\label{sec:rates}

The Galactic ``birthrate'' of long-lived CVs (those with lifetimes of order the Galaxy age) is estimated from observations and population synthesis modeling to be $\mathcal{R} \lesssim 0.01$ yr$^{-1}$ (e.g., \citealt{Ritter&Burkert86,deKool92}).  Thus, if a large fraction $f_{\rm merge}$ of CVs undergo mergers soon after the onset of Roche-lobe overflow instead of living long enough to contribute to the measured CV population (\citealt{Schreiber+16} find $f_{\rm merge} \gtrsim 90\%$), the corresponding rate of the transients described in this paper could be greater by a factor $\sim 1/(1-f_{\rm merger}) \sim 10$ than the CV ``birthrate".  The implied Galactic merger rate of about once per decade is around 10\% of the stellar merger rate inferred from LRN observations ($\gtrsim 0.1$ yr$^{-1}$; \citealt{Kochanek14}) but less than 0.1\% than the rate of classical novae ($\sim 30-70$ yr$^{-1}$; \citealt{Shafter17}).

Given a formation rate of roughly once per decade, if the remnants of such systems shine for $t_{\rm shell} \sim 10^{4}-10^{6}$ yr (Eq.~\ref{eq:tshell}), then we should expect $N_{\rm rem} \sim \mathcal{R}t_{\rm shell} \sim 10^{3}-10^{5}$ such remnants to be present among the Milky Way giant population.

How could such otherwise typical giants ($L \sim 200-5000\, L_{\odot}$, $T_{\rm eff} \approx 3000$ K for $M_{\rm WD} \approx 0.4-0.6M_{\odot}$) be distinguished from those generated by ordinary stellar evolution?   As mentioned in Section \ref{sec:shell}, mixing in the envelope above the burning shell could connect the hot hydrogen shell burning region in CNO equilibrium to the stellar photosphere, thus generating surface abundances atypical for moderately evolved giants.  For example, there exists a rare class of giants which exhibit extreme carbon depletion and lithium over-abundances (e.g., \citealt{Bidelman51,Adamczak&Lambert13,Bond19}).  The large vertical scale-height of the carbon-deficient giants out of the Galactic plane relative to other giants of similar luminosity ($\sim$ apparent stellar age) support them being the ``rejuvenated'' products of binary mass transfer or stellar mergers (e.g., \citealt{Bond19}).  

\subsection{Distinguishing CV Merger Transients from Classical Novae and Stellar Mergers}
\label{sec:distinguish}

The luminosities and timescales of the prompt optical transients which accompany CV mergers (Section \ref{sec:transient}; Fig.~\ref{fig:lightcurve}) are similar to those of classical novae and LRN.  How, then, can one distinguish CV merger events from these much more common transients?

One important difference with respect to novae is the much larger ejecta mass and its solar-like composition; by contrast, nova ejecta are enriched in heavy elements, due to nuclear burning and dredge-up on the WD surface (e.g., \citealt{Gehrz+98}).  Furthermore, the colors of novae generally evolve towards the blue after optical peak, as the hotter WD surface is revealed.  By contrast, CV merger transients will remain red much longer due to obscuration by dust formed in the ejecta (dust also forms in novae but the quantity is much lower).  Nevertheless, some slow classical novae could in principle be misclassified CV mergers (Section \ref{sec:slownovae}).

Regarding LRN, one distinguishing feature with respect to a CV merger is the contrast between the pre- and post-transient luminosities.  In mergers of ordinary (non-degenerate) stars, thermal energy deposited into the primary envelope by the inspiral of the secondary can increase the luminosity of the primary by up to several orders of magnitude relative to its original pre-merger value, particularly when the primary is a low-mass star (e.g., \citealt{Metzger+17,MacLeod+18,Hoadley+20}).  Unfortunately, the resulting luminosity $\sim 10^{2}-10^{3}L_{\odot}$ can reach values comparable to that following a CV merger from residual hydrogen burning on the WD surface, $L_{\rm shell} \gtrsim 10^{3}L_{\odot}$ (Eq.~\ref{eq:Lshell}).  On the other hand, the remnants of LRN will gradually fade in luminosity, on the Kelvin-Helmholtz cooling time of the perturbed stellar envelope (e.g., \citealt{Metzger+17}), a feature which could be detectable in some systems.  Pre-imaging of the transient location (in the Milky Way or nearby galaxies), to characterize the progenitor systems giving rise to intermediate-luminosity transients, could also help identify CV merger candidates.

Another unique feature of CV mergers is the potential for ``precursor" novae.  As discussed in Section \ref{sec:introduction}, drag on the binary from a classical nova eruption may instigate the merger process (e.g., \citealt{Shen15}).  Depending on the relative delay, CV mergers could therefore be accompanied by a nova eruption days to years ahead of the dynamical event we have thus far described.

\begin{table*}
  \begin{center}
    \caption{Galactic Slow Novae}
    \label{tab:table1}
    \begin{tabular}{c|c|c|c|c|c|c|c} 
    \hline
      Name & Eruption & $V_{\rm peak}$ & $t_2$ & $t_3$ & $P_{\rm orb}$ & MS/SG/RG? & Reference \\
 & Year & (mag) & (days) & (days) & (days) &  &  \\
\hline
RR Tel & 1898 & 6.8 & 670 & $>$2000 & & RG-symbiotic	& e.g., \cite{Garcia86}\\
DY Pup & 1902 & 7.0 & 118 & 160 &	0.139 & MS &	\cite{Fuentes-Morales+21} \\
X Ser &	1903 & 8.9 & 400	& &	1.478 & SG & \cite{Thorstensen00} \\
CN Vel & 1905 & 10.2 & 400 & $>$800 & 0.220 & MS &	\cite{Tappert+13}\\
AR Cir & 1906 & 10.5 & 208 & 330 &	0.214 &	MS & \cite{Tappert+13} \\	
V999 Sgr & 1910 & 7.8 &  & 160 &	0.152 & MS & \cite{Mroz+15} \\
BS Sgr & 1917 & 9.2 &	 & 700 &  & MS & \cite{Tappert+15} \\
V849 Oph & 1919 & 7.6 & 140 & 270 & 0.173 & MS & \cite{Shafter+93} \\	
DO Aql & 1925 & 8.5 & 295 & 900 &	0.168 & MS & \cite{Shafter+93} \\
V1310 Sgr & 1935 &	11.7 & & 390 & & ? & This work (Appendix \ref{sec:candidates})\\ 
V356 Aql & 1936 & 7.0 & 127 &	140	& & MS & \cite{DuerbeckSeitter87}\\
BT Mon & 1939 & 8.1 & 118	& 182 &	0.334 & MS & \cite{Smith+98}\\	
V794 Oph & 1939 & 11.7 &	& 220 &	& ?	& \cite{Woudt+03} \\
CT Ser &  1948 & 5 & 100 & & 0.195 & MS & \cite{Ringwald+05}\\	
V365 Car & 1948 & 10.1 &  & 530 & 0.225 & MS & \cite{Tappert+13}\\
V1149 Sgr & 1948 & 7.4 & 	& 210 &	& MS & \cite{Tappert+16}\\
V902 Sco & 1949 & 11.0 &   & 200 & & ?	& This work (Appendix \ref{sec:candidates})\\	
V721 Sco & 1950 & 8.0 &	120	& & & MS/SG &	\cite{Schaefer18}\\
HR Del & 1967 & 3.6 & 167 & 231 & 0.214 & MS & \cite{Kuerster88} \\	
V3645 Sgr & 1970 & 8? & 	 & 300? & &	? & This work (Appendix \ref{sec:candidates})\\

HM Sge & 1975 & 11.0$^a$ & 9700$^a$ & &  & RG-symbiotic & e.g., \cite{Garcia86}\\

V992 Sco & 1992 & 7.7 & 100 & 120 & 0.154 &	MS & \cite{Woudt+03}\\
V723 Cas & 1995 & 7.1 & 263 & 299 &	0.693 & SG & \cite{Goranskij+07}\\	
V445 Pup & 2000 & 8.6 & 215 & 240 &	1.8/3.7 & He star & Steeghs et al.\ in prep\\
V5558 Sgr & 2007 & 8.3 & 281	& 473 &	 & MS & This work (Appendix \ref{sec:candidates}) \\	
\hline
    \end{tabular}
\label{tab:slownovae}
$^{a}$ Photometric parameters for HM Sge's eruption are from \citet{Chochol+04}.
  \end{center}
\end{table*}

\subsection{Constraints on Surviving Companions of Slow Novae}
\label{sec:slownovae}

Given the predicted characteristics of merger transients ($L_{\rm bol} \approx 10^{38}-10^{39}$ erg s$^{-1}$, timescale of $\sim$ months), it is possible that some mergers are hiding amongst the sample of known slow novae. We can observe the host binary after a nova eruption to test that it does indeed remain a binary and that the remaining star is a giant with the expected properties of a merger remnant (Section \ref{sec:shell}). The number of slow novae observed relative to the number of CV mergers (or lack thereof) also constrains how efficiently CAML operates during slow novae.

We used the Galactic nova catalog of \cite{Ozdonmez+18} and took all slow or very slow novae with time to decline from optical maximum by two magnitudes $t_2 > 100$ days (or when $t_2$ is not available, time to decline by three magnitudes $t_3 > 150$ days). There are 27 such novae, but on further study we excluded V1330~Cyg and V5668~Sgr from the sample, as other publications imply relatively high expansion velocities and shorter $t_2/t_3$ \citep{Ciatti&Rosino74, Gordon+21}. 

For the remaining 25 systems, we perused the literature for post-eruption observations of the host system. Table \ref{tab:slownovae} lists the year of nova eruption, peak optical magnitude $V_{\rm peak}$, $t_2$, $t_3$, and the orbital period $P_{\rm orb}$ when available. It also lists constraints on whether the nova host system contains a main sequence (MS), sub-giant (SG), red giant (RG; including asymptotic giant branch), or helium star, and the reference to this constraint and the orbital period. Fourteen have well-studied binaries with orbital periods measured after outburst (see Table \ref{tab:slownovae}); these are novae that certainly did not end in a merger. One of these is the unique V445 Pup, which is the only Helium nova ever observed and likely samples a different evolutionary path than the other H novae studied here \citep{Ashok&Banerjee03, Woudt+09}. Two other systems (RR~Tel and HM~Sge) are well-observed symbiotic stars with Mira giant companions; they show evidence for a WD coexisting with the evolved star, and anyway are not candidates for the CV merger scenario because the orbital separation is so great that CAML should not be able to act.

For the remaining 9 systems, we can use photometric and spectroscopic measurements to constrain the nature of the system after eruption and differentiate a giant from a dwarf or mildly evolved donor (see works by \citealt{Weight+94, Darnley+12, Pagnotta&Schaefer14}).  
Three of these systems (BS~Sgr, V356~Aql, V1149~Sgr) have spectra showing emission lines indicative of a CV and no evidence of an evolved companion \citep{DuerbeckSeitter87,Ringwald+96,Tappert+14,Tappert+15,Tappert+16}. 
In the other six cases, very little is known, or there are published indications of a bright giant-like star at the position of the nova.  

We investigate these six old novae in more detail in Appendix \ref{sec:candidates}.  In two cases (V5558 Sgr and V721 Sco), we find evidence that any putative companion would be low luminosity and unevolved, inconsistent with a merger remnant.  However, in the other four cases (V1310 Sgr, V794 Oph, V3645 Sgr, V902 Sco), we find that a merger candidate cannot be absolutely excluded. The astrometric positions of the novae are poorly known, leading to ambiguity in identifying the post-nova system.

In summary, out of 22 H slow classical novae known in our Galaxy (excluding V445 Pup, RR~Tel, and HM~Sgr), there are four cases where we can not completely exclude the behavior expected following a CV merger: bloating to a giant state (Section \ref{sec:shell}), with no evidence for a hot WD companion.  These four systems are excellent targets in search of signatures of ongoing accretion (e.g., optical emission lines, X-rays).  The implication from the population of slow novae is that $\lesssim 4/22 \sim 18\%$ of slow novae end in merger.  To the extent that the Galactic slow nova population is complete over the past century (within a factor of $\lesssim$ few; \citealt{Kawash+21}), this implies a Galactic CV merger rate of less than one per decade, broadly consistent (albeit with large uncertainties) with the rate predictions in Section \ref{sec:rates}.

\subsection{Formation of Isolated Low-Mass WDs}

Low-mass WDs with $M_{\rm WD} \lesssim 0.5M_{\odot}$ cannot descend from single stars because their formation time is longer than the age of the Universe.  Indeed, the vast majority of low mass WDs belong to close binary systems (e.g., \citealt{Marsh+95,Brown+10}).  However, a significant fraction $\lesssim 20-30\%$ of these appear to be single (e.g., \citealt{Brown+11}).  Many explanations have been forwarded to explain low-mass single WDs, ranging from common envelope events (e.g., \citealt{Nelemans10}) to strong mass-loss in metal-rich stars (e.g., \citealt{Kilic+07}) to the remnant companions of Type Ia supernovae (e.g., \citealt{Justham+09}).  

\citet{Zorotovic&Schreiber17} found that the CV merger rate predicted by CAML is consistent with that required to explain the low-mass WD population.  Insofar that we predict that a large fraction $\gtrsim 90\%$ of the companion is ejected in disk outflows instead of being accreted (Eq.~\ref{eq:Macc}; Fig.~\ref{fig:allmass}), we confirm that the end product of the merger of a low mass WD with its CV companion will remain  a low mass WD.  We note, however, even in the limiting case where the WD accretes all of the secondary star, \citet{Zorotovic&Schreiber17} still found that CV mergers could create a separate population of WDs with masses extending below the population generated by single-star evolution.    

\subsection{CK Vul}
\label{sec:CKVul}

The transient giving rise to CK Vul was observed in the years 1670-1672, but no counterpart was identified until a bipolar nebula was detected at its location (e.g., \citealt{Shara+82,Shara+85}).  \citet{Kato03} and \citet{Kaminski+15} proposed that CK Vul was a stellar merger event.  \citet{Eyres+18} instead favor a merger involving a WD and brown dwarf, based (in part) on the presence of ionized species HCO$^{+}$ and N$_{2}$H$^{+}$, which require exposure to an intense UV radiation field, such as that supplied by a hot central WD.  However, \citet{Kaminski+15,Kaminski+20a,Kaminski+20b} instead argue that shock excitation by outflows from the remnant can generate the emission lines without appealing to an independent source of ionizing photons.

CK Vul is surrounded by chemically rich molecular gas with non-solar isotopic ratios, unlike those found in the ejecta of classical novae.  Estimates for the total gaseous mass in the nebula are $\gtrsim 0.6M_{\odot}$ \citep{Kaminski+15,Banerjee+20}.  This range is again too large for a classical nova, but potentially consistent with a stellar merger event \citep{Kato03,Kaminski+15} or a merger between a WD and a low-mass stellar object \citep{Eyres+18}.

Was CK Vul a CV merger of the type described in this paper?  The present-day luminosity of the remnant $\sim 20 L_{\odot}$ \citep{Hajduk+07,Eyres+18,Banerjee+20} is orders of magnitude weaker than predicted if the remnant were still undergoing hydrogen shell burning (Eq.~\ref{eq:Lshell}).  However, this may not be an irreconcilable problem for this scenario; if the accreted mass is sufficiently small $\lesssim 10^{-4}M_{\odot}$ and/or the WD sufficiently massive $\gtrsim M_{\odot}$, the shell burning lifetime $t_{\rm shell}$ (Eq.~\ref{eq:tshell}) could in principle be less than the present age of CK Vul of $\lesssim 350$ yr, consistent with its present (much lower) luminosity being that of a young (``refreshed'') WD.

Interestingly, the remnant of CK Vul exhibits an over-abundance of $^{7}$Li \citep{Hajduk+07}, indicative of nuclear processing of the ejecta (e.g., \citealt{Cameron&Fowler71}).  The molecule $^{26}$AlF is also detected in the remnant of CK Vul \citep{Kaminski+18}, the radioactive $^{26}$Al isotope of which was likely produced via the $^{25}$Mg(p,$\gamma$)$^{26}$Al reaction at high temperature $\gtrsim 3\times 10^{7}$ K.  The conditions required to generate $^{7}$Li and $^{26}$Al could also in principle be achieved in the burning layer of a massive WD, which if mixed to the photosphere during the shell burning phase and released in an outflow could pollute the nebula (Section \ref{sec:shell}).\footnote{\citet{Kaminski+18} attribute the processed material to matter dredged up and ejecta from the vicinity of the helium core of an evolved progenitor during a stellar merger.}  

On the other hand, a CV hosting a massive WD is not the most natural system to undergo a merger under the CAML evolutionary paradigm.  Another mechanism to instigate the merger of a star or brown dwarf with the WD, such as dynamical interaction in a triple system, could instead be favored.  In a scenario where a tertiary star is involved in driving an inner binary to merge (e.g., via the Kozai-Lidov mechanism), then the multiple light curve peaks in CK Vul over the course of years, could be produced by a gradual sequence of grazing encounters (``partial tidal disruptions''), prior to the final complete disruption event.  A broadly similar model has been outlined to explain the Giant eruption from Eta Car and the formation of its nebula (e.g., \citealt{Hirai+21} and references therein).

The nebula surrounding the remnant of CK Vul is bipolar in shape and extends to a radius $\sim 0.5$ pc.  Given the known age of the source, the material responsible for generating these bipolar lobes must be expanding at a velocity of several hundred km s$^{-1}$.  Although the morphology of the bipolar ejecta exhibit similarities to the polar wind ejecta seen in the final snapshot of our simulations (Fig.~\ref{fig:snapshots}), the average velocity we predict $\sim 800$ km s$^{-1}$ (Fig.~\ref{fig:vel_distribution}) is roughly consistent with those observed.  On the other hand, \citet{Banerjee+20} infer velocities up to $\sim 2100$ km s$^{-1}$ at the tips of the bipolar lobes in CK Vul, higher than the maximum velocities found in our simulations. 

Another feature of the CK Vul remnant is a warped dusty molecule-rich disk which extends to radial scales $\gtrsim 10^{16}$ cm from the central remnant and exhibits its own bipolar wind/outflow \citep{Kaminski+15,Eyres+18}.  As discussed at the end of Section \ref{sec:analytic}, the disk formed from the disrupted secondary in a CV merger will continue to expand to large radii as the result of outwards distribution of angular momentum by viscosity (see the final snapshot in Fig.~\ref{fig:rotational_support}).  The expanding disk will continue to cool to the point that dust and molecule formation is possible (see \citealt{Margalit&Metzger17} for a discussion of the long-term torus evolution in the context of WD-neutron star mergers).  However, the size of the ``disk'' surrounding CK Vul is probably too large to be a bound hydrostatic remnant from a CV merger.  Nevertheless, we speculate that the observed disk-like structure may instead be a slow unbound outflow from a smaller disk, with a velocity $\sim 10$ km s$^{-1}$ driven by photo-ionization heating from the hot central WD, similar to those observed around massive proto-stars (e.g., \citealt{Hollenbach94}).  As suggested by \citet{Eyres+18}, the observed warping and precession of the disk could also be driven by irradiation (e.g., \citealt{Nixon&Pringle10}).

\section{Conclusions}
\label{sec:conclusion}

The conclusions of our study can be summarized as follows:
\begin{itemize}

\item{Recent observational and theoretical studies suggest a scenario in which a large fraction of CV binaries, particularly those harboring low mass WDs, are being prematurely ``removed'' from the CV population.  One way this could occur is through the onset of unstable mass-transfer, driven by angular momentum loss during or following classical novae, resulting in a CV merger.  Regardless of whether the CAML mechanism or some other process is at work, being able to directly observe or constrain the occurrence of these events is of interest in interpreting the CV population.  }

\item{The process of unstable mass transfer culminates in the dynamical disruption of the secondary into a massive disk surrounding the WD, which subsequently accretes over the course of days at highly super-Eddington rates.  Hydrodynamical $\alpha$-viscosity simulations of the accretion phase reveal that outflows unbind a large fraction $\gtrsim 90\%$ of the secondary mass, resulting in the ejection of a mass $\gtrsim 0.1M_{\odot}$ at characteristic velocities $\sim 500-1000$ km s$^{-1}$.  Insofar that the WD mass does not grow appreciably during this process, we concur with previous work (\citealt{Zorotovic&Schreiber17}) that CV mergers offer a channel for generating single low-mass WDs.}

\item{Radiation released in the expanding disk ejecta, primarily energized by hydrogen recombination, powers optical transient emission with a peak luminosity $L_{\rm pk} \sim 10^{38}-10^{39}$ erg s$^{-1}$ and characteristic timescale of a couple months.  The predicted luminosities/timescales of CV merger transients overlap those of slow classical novae and luminous red novae from ordinary (non-degenerate) stellar mergers.  Soon after the optical peak, the ejecta shell will form copious amounts of dust and molecules, enshrouding the merger remnant for decades after the merger.}

\item{Over a similar timescale of decades, the mass remaining on the WD surface will undergo hydrogen shell burning, inflating the remnant into a giant star of luminosity $\sim 300-5000L_{\odot}$, effective temperature $T_{\rm eff} \approx 3000$ K, and lifetime $\approx 10^{3}-10^{5}$ yr.  These remnants will thus constitute a modest population of giants in the Milky Way, possibly characterized by atypical abundances (e.g., CNO nuclei or $^{7}$Li) due to rotational or convective mixing bringing shell-burning products to the surface.}    

\item{Given the estimated rate of CV mergers and their observational appearance similar to slow novae, we could expect as many as $\sim 10$ slow novae per century to in fact be masquerading CV mergers.  Of the 22 systems that are photometrically or spectroscopically constrained to be H slow classical novae over the last $\sim$century (Table \ref{tab:slownovae}), we identify four systems (V1310 Sgr, V794 Oph, V3645 Sgr, and V902 Sco) for which a giant star merger remnant cannot be ruled out by the photometry of the noca system (Appendix \ref{sec:candidates}).  We encourage observational follow-up of these sources to look for signatures of a surviving WD companion or, alternatively, evidence in support of a merger.}

\item{The historical transient CK Vul was recently suggested to be a merger between a WD and brown dwarf \citep{Eyres+18}.  We find some similarities between the morphology of the molecular nebula and those predicted by our hydrodynamical simulations of CV mergers, such as the presence of a bipolar outflow and an extended disk-like structure.  However, reproducing the low luminosity of the remnant at the current epoch, and the nebula abundance anomalies (particularly $^{26}$Al), probably require a high mass WD primary, inconsistent with a CV merger driven by the empirical CAML model of \citet{Schreiber+16}.  Nevertheless, other mechanisms may be capable of driving a hydrogen-rich star to undergo a binary merge with a WD (e.g., dynamical interactions involving a third body) that could result in a qualitatively similar outcome to the systems studied here.}

\end{itemize}

\acknowledgements
We are grateful for helpful conversations and observing assistance of E.\ Aydi, A.\ Kawash, and K.\ Sokolovsky.
We are grateful for helpful information from M.\ Schreiber, N.\ Evans, and M.\ Zorotovic.  BDM acknowledge support from NSF grant AST-2009255.  LC acknowledges support from NSF grant AST-1751874. KJS acknowledges support from NASA through the Astrophysics Theory Program (NNX17AG28G \& 80NSSC20K0544).  JS acknowledges support from the Packard Foundation. This work has made use of data from the European Space Agency (ESA) mission
{\it Gaia} (\url{https://www.cosmos.esa.int/gaia}), processed by the {\it Gaia}
Data Processing and Analysis Consortium (DPAC,
\url{https://www.cosmos.esa.int/web/gaia/dpac/consortium}). Funding for the DPAC
has been provided by national institutions, in particular the institutions
participating in the {\it Gaia} Multilateral Agreement.

\appendix

\section{1D Steady-State Disk Model}
\label{sec:1D}

Here we construct steady-state one-dimensional (height-integrated) inflow models across radii from $R_{\rm in} = R_{\rm WD}$ to $R_{\rm out} = R_{\rm d,0}$, following the procedure outlined in \citet{Metzger12} and \citet{Margalit&Metzger16}.

The mass accretion rate obeys
\be
\dot{M} = \dot{M}_{0}\left(\frac{r}{R_{\rm d,0}}\right)^{p} = 3\pi \nu \Sigma = 3\pi \alpha r^{2}\Omega_{\rm K}\theta^{2}\Sigma,
\label{eq:Sigma}
\ee
while the midplane density obeys
\be
\rho = \frac{\Sigma}{2\theta r} = \frac{\dot{M}}{6\pi \alpha r^{3}\Omega_{\rm K} \theta^{2}},
\ee
where 
\be
\Omega_{\rm K} \simeq \left(\frac{G\left[M_{\rm WD}+M_{\star}\right]}{r^{3}}\right)^{1/2}
\ee
The steady-state disk aspect ratio can be written (Margalit \& Metzger 2016; their Eq.~28)
\be
\theta \simeq \sqrt{\frac{\gamma-1}{2\gamma}(1+2\mathcal{B}e'_{\rm crit})} \underset{\gamma = 5/3, \mathcal{B}e'_{\rm crit} = 0}\approx 0.45,
\ee
where $\gamma \simeq 5/3$ is the adiabatic index and $\mathcal{B}e'_{\rm crit} \lesssim 0$ is the Bernoulli parameter to which the disk is regulated via disk outflow cooling.

\begin{figure}
    \centering
    \includegraphics[width=0.45\textwidth]{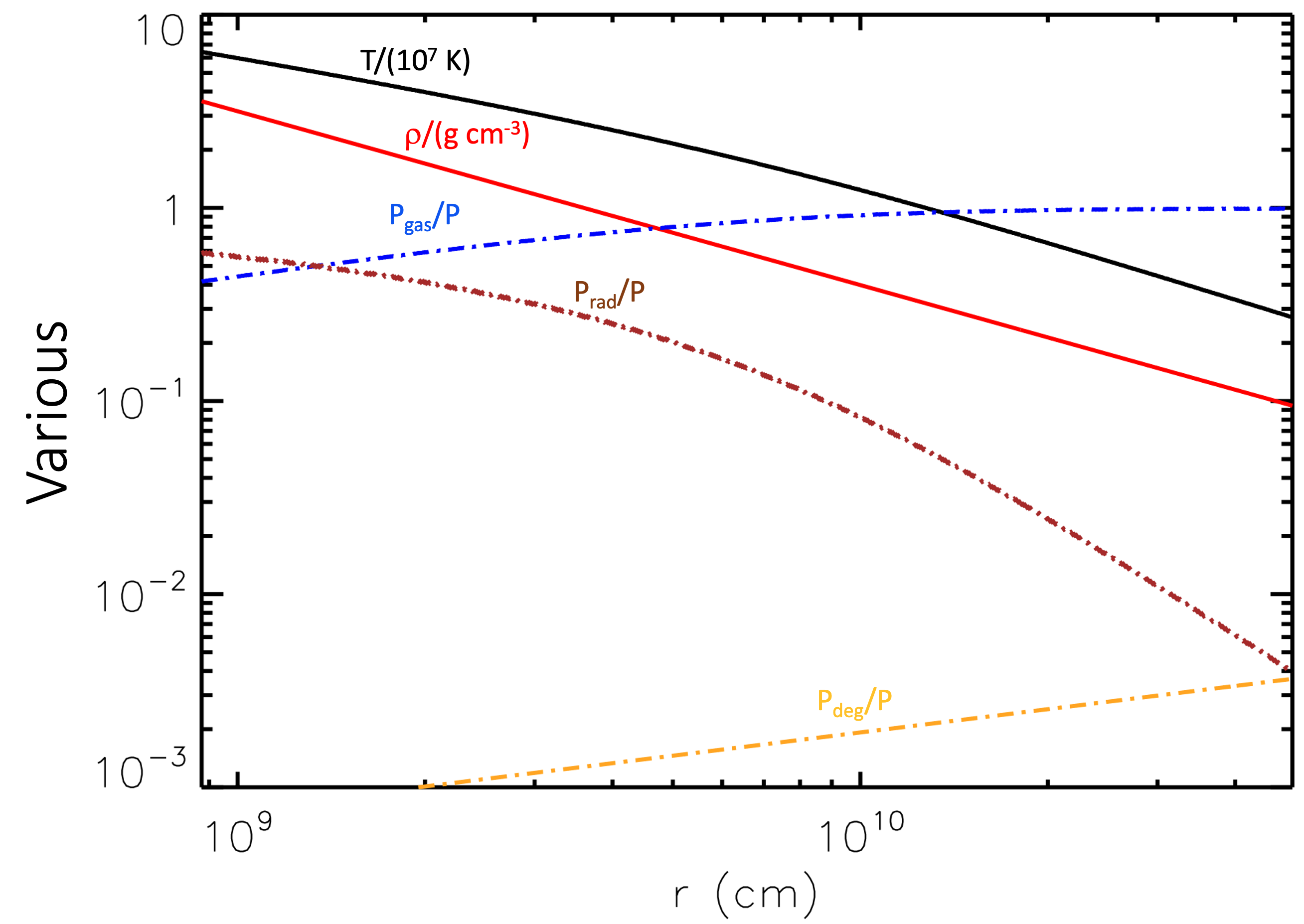}
    \includegraphics[width=0.45\textwidth]{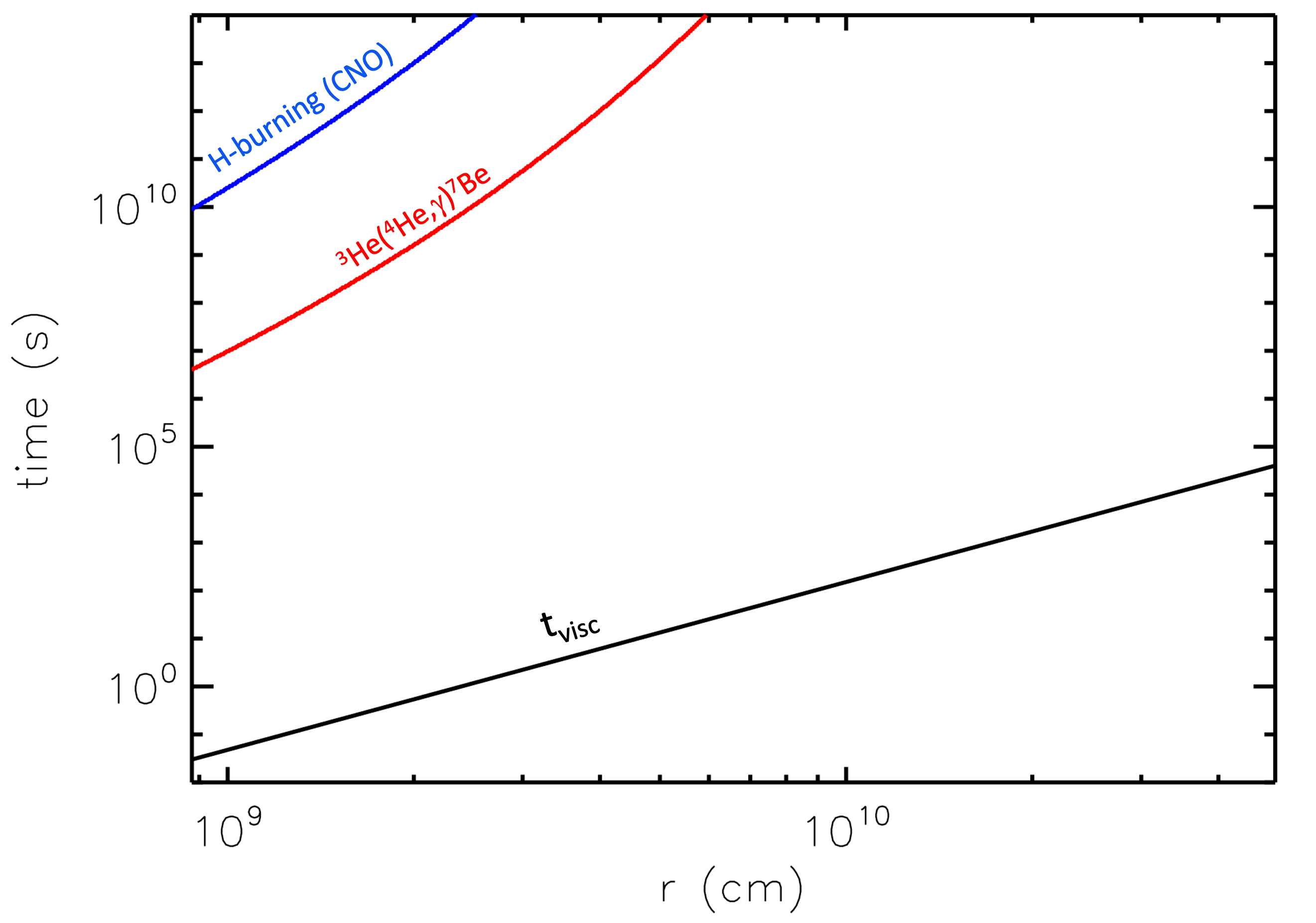}
    \caption{One-dimensional steady state model of an accretion disk of 
    mass $M_{\rm d,0}=$ 0.2$M_{\odot}$ onto a $M_{\rm WD} = 0.6M_{\odot}$ WD primary.  We have assumed a viscosity $\alpha = 0.1$ and wind mass-loss parameter $p = 0.6$ (Eq.~\ref{eq:Sigma}). {\bf Left:} Various thermodynamic quantities, including the midplane temperature $T$ (black), density $\rho$ (red), and fractional contributions to the total pressure $P$ from ideal gas (blue), radiation (brown), and electron degeneracy (orange).  {\bf Right:} Viscous/inflow timescale (black) compared to nuclear burning times of a few key reactions (colored).  The nuclear burning times are much longer than the burning times at all radii in the disk, demonstrating that no appreciable burning is expected during the accretion phase after the merger.}
    \label{fig:burning}
\end{figure}

The radial temperature profile then follows from,
\be
\theta = \frac{2P}{\Sigma r \Omega_{\rm K}^{2}} \Rightarrow P = \frac{\dot{M}\Omega_{\rm K}}{6\pi \alpha r \theta} 
\ee
where the total midplane pressure
\be
P(\rho,T) = P_{\rm gas} + P_{\rm rad} + P_{\rm deg}
\ee
\be
P_{\rm gas} = \frac{\rho kT}{\mu m_p}; \,\,\,\, P_{\rm rad} = \frac{a}{3}T^{4}; \,\,\,\,P_{\rm deg} = \frac{h^{2}}{20 m_e}\left(\frac{3}{\pi}\right)^{2/3}\left(\frac{\rho}{\mu_e m_p}\right)^{5/3} 
\ee
The left panel of Figure \ref{fig:burning} shows radial profiles for an example 1D steady-state disk solution for the fiducial case of a secondary of mass $M_{\star} = 0.2M_{\odot}$ accreting onto a 0.6$M_{\odot}$ WD with a viscosity $\alpha = 0.1$ and wind mass-loss parameter $p = 0.6$.  The right panel of Fig.~\ref{fig:burning} compares the local viscous timescale of the disk (Eq.~\ref{eq:tvisclocal}) to the timescale for nuclear burning, $t_{\rm nuc}$, for reactions relevant to hydrogen burning and lithium production.  The fact that $t_{\rm visc} \ll t_{\rm nuc}$ at all radii shows that nuclear reactions can be neglected during the disk evolution and the disk outflows will possess close to the original composition of the secondary star (Section \ref{sec:burning}).

\section{Light Curve Model}
\label{sec:lightcurve}

Here, we present a semi-analytic model for the optical light curves of CV merger transients, or more generally events powered by the expansion of hot ionized material.  We begin with a one-zone model of a uniform shell expanding with a single velocity, which we then expand to a multi-zone calculation that accounts for the more realistic case of ejecta with a range of velocities.

\subsection{One-Zone Model}
 We consider the ejecta shell to be spherical and of mass $M_{\rm ej}$, velocity $v_{\rm ej}$, and thermal energy $E$.  Gas pressure is assumed to dominate throughout the evolution (Figure \ref{fig:snapshots}), in which case the internal energy is related to the average temperature of the ejecta $T$ according to $E \simeq 3/2(M_{\rm ej}/\mu m_p)kT$, where $\mu$ is the mean molecular weight.  The internal energy evolves with time $t$ since release of the shell, according to
\be
\frac{dE}{dt} = -\frac{3(\gamma_3 -1)E_{\rm rad}}{R_{\rm ej}}v_{\rm ej} - L_{\rm rad},
\label{eq:Edot}
\ee
where $R_{\rm ej} = v_{\rm ej}t$ is the ejecta radius.  The first term in Eq.~(\ref{eq:Edot}) accounts for $PdV$ losses, where $\gamma_{3}$ is the effective adiabatic index (see below).  The second term,
\be
L_{\rm rad} = \frac{E_{\rm rad}}{t_{\rm d} + t_{\rm LC}},
\ee 
accounts for radiative losses, where $E_{\rm rad} = a T^{4}V_{\rm ej}$ is the radiation energy, $V_{\rm ej} = 4\pi R_{\rm ej}^{3}/3$, and 
\be
t_{\rm d}  = \frac{M_{\rm ej}\kappa}{4\pi c R_{\rm ej}}
\ee 
is the photon diffusion time, $\kappa$ is an Rosseland opacity, and $t_{\rm LC} = R_{\rm ej}/c$ the ejecta light crossing time (included for numerical stability as a physical lower limit on the timescale of thermal energy release).  For the opacity we use the approximate analytic formula (e.g., \citealt{Metzger&Pejcha17}) for solar metallicity gas (metallicity $Z = 0.02$, hydrogen mass-fraction $X = 0.74$):
\be
\kappa(\rho,T) = \kappa_{\rm m} + \left(\kappa_{\rm H^{-}}^{-1} + (\kappa_{\rm e}+\kappa_{\rm K})^{-1}\right)^{-1} ,
\ee 
which accounts for electron scattering $\kappa_{\rm e} \approx 0.38 X x_{\rm ion}$ cm$^{2}$ g$^{-1}$, bound-free/free-free absorption $\kappa_{\rm K} \approx 4\times 10^{25}Z(1+X)\rho T^{-7/2}$ cm$^{2}$ g$^{-1}$, H$^{-}$ opacity $\kappa_{\rm H^{-}} \approx 1.1\times 10^{-25}Z^{0.5}\rho^{0.5}T^{7.7}$ cm$^{2}$ g$^{-1}$ and a characteristic molecular opacity $\kappa_{\rm m} \approx 0.1 Z$ cm$^{2}$ g$^{-1}$, where $\rho = 3M_{\rm ej}/(4\pi R_{\rm ej}^{3})$ is the ejecta density.  Radiation is trapped in the flow until after helium has recombined, so we only consider the ionization of hydrogen in calculating the opacity and other ejecta properties.

The hydrogen ionization fraction $x_{\rm ion}$ is calculated from the Saha equation according to,
\be
\frac{x_{\rm ion}^{2}}{1-x_{\rm ion}} = \frac{m_p}{X\rho}\left(\frac{2\pi m_e kT}{h^{2}}\right)^{3/2}\exp\left[-\frac{\rm E_{\rm Ryd}}{kT}\right],
\ee 
where $E_{\rm Ryd} = 13.6$ eV. 

Neglecting radiation pressure, the effective adiabatic index, including the effects of hydrogen recombination, can be written (e.g., \citealt{Kasen&RamirezRuiz10})
\be
\gamma_3-1 = \frac{(1+\bar{x}) + X \frac{x_{\rm ion}(1-x_{\rm ion})}{(2-x_{\rm ion})}\left(\frac{3}{2}+\frac{E_{\rm Ryd}}{kT}\right)}{\frac{3}{2}(1+ \bar{x}) + X\frac{x_{\rm ion}(1-x_{\rm ion})}{(2-x_{\rm ion})}\left(\frac{3}{2}+\frac{E_{\rm Ryd}}{kT}\right)^{2}},
\ee
where $\bar{x} \equiv Xx_{\rm ion}$.  We note that in the limit of neutral ($x_{\rm ion} \approx 0$) or fully ionized gas  ($x_{\rm ion} \approx 1$), we obtain $\gamma_{3} = 5/3$, the usual adiabatic index of a monotonic gas.  However, during recombination ($x_{\rm ion} \approx 0.5$), $\gamma_{3}-1$ can be small, corresponding to the nearly isothermal evolution enforced by the energy released by hydrogen recombination.  Again, we neglect helium recombination.

As initial conditions, we assume take the ejecta radius to be $R_{\rm d,0}$ at time $t_0 = R_{\rm d,0}/v_{\rm ej}$.  The initial thermal energy is given by,
\be
E(t_0) \equiv E_0 = \frac{3}{2}\frac{M_{\rm ej}}{\mu m_p}kT_0
\ee
Once matter has been ejected from the disk and is no longer being heated by viscosity, its specific entropy $s \propto ln(T^{3/2}/\rho)$ will remain constant (until radiation losses become important).  Introducing an entropy-like quantity,
\be
\mathcal{S} \equiv \frac{T}{\rho^{2/3}} = constant,
\label{eq:S}
\ee
we can write
\be
T_0 = \mathcal{S}\rho_0^{2/3} = \mathcal{S}\left(\frac{3M_{\rm ej}}{4\pi R_{\rm d,0}^{3}}\right)^{2/3}
\ee
We thus define a dimensionless initial energy,
\begin{eqnarray}
\eta \equiv \frac{E_0}{M_{\rm ej}v_{\rm ej}^{2}} = \frac{\mathcal{S}}{v_{\rm ej}^{2}}\frac{3}{2}\frac{k}{\mu m_p}\left(\frac{3M_{\rm ej}}{4\pi R_{\rm d,0}^{3}}\right)^{2/3} 
\approx 120\left(\frac{\mathcal{S}/v_{\rm ej}^{2}}{10^{-6}\rm K\,s^{2}\,g^{-2/3}}\right)\left(\frac{R_{\rm d,0}}{R_{\odot}}\right)^{-2}\left(\frac{M_{\rm ej}}{0.3M_{\odot}}\right)^{2/3}
\label{eq:eta}
\end{eqnarray}
This expression allows us to relate the value of $\mathcal{S}/v_{\rm ej}^{2}$ measured in our simulation data (once matter has been unbound from the disk), to determine the values of $\eta$ which serve as input to the light curve calculations.  For values of $R_{\rm d,0}$ comparable to the outer disk radius (Eq.~\ref{eq:Rc}), we find $\mathcal{S}/v_{\rm ej}^{2} \sim 10^{-7}$ K s$^{2}$ g$^{-2/3}$ and hence $\eta \sim 100$ using our simulation output (top panel of Fig.~\ref{fig:vel_distribution}).

The dashed lines in Figure \ref{fig:lightcurve} shows results from one-zone model, calculated for $M_{\rm ej} = 0.4M_{\odot}$, $R_{\rm d,0} = R_{\odot}$, $\eta = 100$, and for different values of $v_{\rm ej} = 400, 700, 1000$ km s$^{-1}$.  The one-zone light curves rise to an extremely sharp peak at a luminosity $\approx 10^{40}$ erg s$^{-1}$ on a timescale that scales inversely with $v_{\rm ej}$, as expected from diffusion timescale arguments.  The sharp light curve peak is driven by the sudden opacity drop that occurs as the ejecta recombines.  While this model reasonably captures the total energy release from hydrogen recombination, the rapid light curve evolution is not physical because the one-zone model misses the finite amount of time for the cooling wave to propagate back through the ejecta shell.  This deficiency is removed by considering a multi-zone model.  

\subsection{Multi-Zone Model}

We now consider the ejecta to possess a distribution of velocities $v_{\rm ej}$, such that enclosed mass above a given velocity given by 
\be
M_{\rm ej}(>v_{\rm ej}) = \int_{v_{\rm ej}}^{\infty}\frac{dM}{dv}dv \simeq M_{\rm ej,tot}e^{-(v_{\rm ej}/\bar{v}_{\rm ej})^{3}},
\label{eq:Menc}
\ee
where $M_{\rm ej,tot}$ is the total ejecta mass and $\bar{v}_{\rm ej}$ a free parameter.  For example, our numerical simulations (Fig.~\ref{fig:vel_distribution}) motivate this functional form, with $M_{\rm ej,tot} \simeq 0.2M_{\odot}$ and $\bar{v}_{\rm ej} \approx 700$ km s$^{-1}$ in the fudicial model.  

As each shell $dM$ becomes transparent, it will contribute to the total luminosity, starting with the highest velocity material (outermost layers).  The total luminosity is thus given by summing the contribution from each shell,
\be
L_{\rm rad,tot}(t) = \frac{1}{M_{\rm ej,tot}}\int_{0}^{\infty}\frac{dM}{dv}L_{\rm rad}\left[t,M_{\rm ej} = M_{\rm ej}(>v),v_{\rm ej}=v\right]dv,
\ee
where $L_{\rm rad}(t,M_{\rm ej},v_{\rm ej})$ is calculated using the one-zone model as in the previous section.  The solid line in Fig.~\ref{fig:lightcurve} shows the multi-zone model using Eq.~(\ref{eq:Menc}) and a constant value $\eta = 100$ for all mass-shells.

\section{Candidate Merger Remnants from Galactic Slow Novae}
\label{sec:candidates}

\subsection{V1310 Sgr (1935)}

V1310 Sgr has a claimed red giant counterpart \citep{Pagnotta&Schaefer14}, based on a tentative association of the poorly observed eruption with a bright star in quiescence \citep{Downes+01}. \citet{Tappert+12} obtained a spectrum and indeed confirmed the object as a Mira giant. However, \citet{Tappert+14} acknowledge some uncertainty in the association of the Mira giant with the 1935 nova eruption. We analyzed the finder chart of \citet{Fokker51}, and indeed find that the Mira giant is offset southeast of the nova position by about 20$^{\prime\prime}$. The revised best FK5 position is RA = 18h35m00.50s, Dec = $-30^{\circ}03^{\prime}14.1^{\prime\prime}$, with an estimated uncertainty of $2-3^{\prime\prime}$. The only bright star ($g \lesssim 20$ mag) within this region has Pan-STARRS $g = 18.1$ mag \citep{Chambers+16}, VVV $J = 15.7$ mag \citep{Minniti+17}, and is located at RA = 18h35m00.480s, Dec = $-30^{\circ}03^{\prime}17.24^{\prime\prime}$. We obtained a spectrum of this star on 2021 Jul 3.5 using the Goodman High Throughput Spectrograph \citep{Clemens+04} on the 4.1\,m Southern Astrophysical Research (SOAR) telescope, making use of the 400~l\,mm$^{-1}$ grating and covering 4000--7910\,\AA.
The spectrum was reduced and optimally extracted using standard tools in IRAF \citep{Tody86}. The spectrum is indicative of a cool star, moderately reddened, and the strongest absorption lines are of Mg and the infrared Ca triplet; there is no evidence of emission lines in the spectrum. If the star is located in the Galactic bulge $\sim$8 kpc away (its \emph{Gaia} EDR3 parallax is $-0.03\pm0.12$ mas), its $J$-band absolute magnitude would be $M_J = 1.1$ mag, as expected for dwarf or mildly evolved companions in nova host systems \citep{Darnley+12}.
Based on all the information in hand, the star could be associated with V1310 Sgr, but there is no evidence of an accreting WD.  If the star marks the product of a CV merger, WD signatures would be expected to be absent, but we would also expect a more luminous giant. However, it is also quite possible that the star we obtained a spectrum of is merely a chance interloper, and the actual star associated with V1310 Sgr is substantially fainter. Clearly, high-quality astrometry of slow novae in eruption is needed, to enable accurate study of these systems in quiescence.

\subsection{V794 Oph (1939)}

V794 Oph was a poorly observed slow nova that erupted in 1939. It has been claimed to be associated with a giant star \citep{Duerbeck88,Pagnotta&Schaefer14} based on coordinates published in the \citet{Duerbeck87} catalog of novae. However, these coordinates are quite speculative, attempting to hone the position originally published by \citet{Burwell&Hoffleit43} based on objective prism plates and only quoted to arcminute precision. There is a relatively bright star at the  \citet{Duerbeck87} position ($\sim$17.7 mag) which would imply a low amplitude for the nova eruption ($\sim$6 mag) and therefore an association with a giant \citep{Duerbeck88}. \citet{Ringwald+96} obtained a spectrum of this star and observe red continuum with some faint absorption features; \citet{Surina14} finds that the equivalent width of Ca\,I features are indicative of a giant star. However, \citet{Woudt+03} observed this giant for 1--2 hr on two separate occasions and find no evidence of variability, which makes it unlikely that this giant is in fact associated with an accreting WD (although we note that it \emph{could} be consistent with a stable giant remnant of a CV merger).  \citet{Woudt+03} point out that there are no other variable sources in their field consistent with the position of \citet{Burwell&Hoffleit43}. It seems likely that the giant star was not associated with the eruption of V794~Oph, and V794~Oph is a much fainter and less conspicuous source. However, the presence of a giant near V794~Oph's position means we can not exclude the possibility that V794~Oph's eruption marked a CV merger.

\subsection{V902 Sco (1949)}

V902 Sco has essentially nothing published about its post-eruption state.  As \citet{Duerbeck87} notes, the position and quiescent source association are uncertain, as the only published finding chart for V902 Sco is an objective prism image which naturally offers poor astrometry \citep{Henize61}, and the  nova erupted in a crowded and heavily extinguished region of the Galaxy.
We obtained SOAR spectra (with the same setup as for V1310 Sgr, on 2021 Jul 6.0) of the two stars suggested by \citet{Duerbeck87} as potential  quiescent counterparts  to V902 Sco. The first star has DECaPS $g =21.40$ mag \citep{Schlafly+18}, VVV $J= 15.1$ mag \citep{Minniti+17}, and is located at ICRS coordinates RA = 17h26m08.393s, Dec = $-39^{\circ}04^{\prime}02.66^{\prime\prime}$. The second has $g =20.49$ mag, $J=16.0$ mag, and is located at \emph{Gaia} ICRS coordinates RA = 17h26m08.552s, Dec = $-39^{\circ}04^{\prime}06.44^{\prime\prime}$. The spectra of both objects are consistent with  cool, dust-reddened stars, with no sign of emission lines in either case. Neither star has a significant \emph{Gaia} parallax measured in EDR3 (parallaxes are $-0.08\pm0.27$ mas and $0.44\pm0.33$ mas, respectively; \citealt{Gaia16, Gaia21}). The extinction along this line of sight yields $A_g = 4.6$ mag ($A_J= 1.0$ mag; \citealt{Schlafly+11}). Assuming that V902~Sco is located in the Galactic bulge $\sim$8 kpc away and is behind the full extinction column, the stars would have J-band absolute magnitudes of $M_J = 0.5$ and  $M_J = -0.4$ mag, respectively; they could be even more luminous if they were located further away. Based on this analysis alone, we can not rule out that one of these stars is evolved 
\citep{Darnley+12}, and is associated with V902 Sco.  Therefore, V902~Sco remains a candidate for a CV merger. 

\subsection{V721 Sco (1950)}

Although little is known about the 1950 eruption of V721~Sco, \citet{Duerbeck87} claims a secure association with a quiescent source located at ICRS RA = 17h42m29.095s, Dec $= -34^{\circ}40^{\prime} 41.53^{\prime\prime}$ and with measured brightnesses $g = 16.3$ mag in DECaPS, $J= 13.6$ mag in VVV. The source also has a secure \emph{Gaia} parallax measurement of $0.54\pm0.04$ mas in EDR3, implying a relatively nearby distance of $\sim$1.5 kpc \citep{Schaefer18}. Conservatively assuming this source is behind the full absorbing column along that line of sight ($A_g = 3.6$ mag; \citealt{Schlafly+11}) yields absolute magnitude estimates, $M_g =1.8$ mag and $M_J = 1.9$ mag, which are indicative of a main sequence or subgiant counterpart \citep{Darnley+12}.

\subsection{V3645 Sgr (1970)}

V3645 Sgr erupted in 1970 but was also poorly observed. \citet{Sarajedini84} measured its position on plates from the Maria Mitchell Observatory in outburst; there is a faint star at this location, measured at $g = 20.1$ mag in Pan-STARRS and undetected in 2MASS ($J \gtrsim 16$ mag). These same plates were measured by \citet{Duerbeck87}, who found a disparate position 11$^{\prime\prime}$ to the northwest (but acknowledge some uncertainty in this crowded field) matched to a somewhat brighter star (Pan-STARRS $g = 18.7$ mag, 2MASS $J = 15.5$ mag). Based on the position of the \citet{Duerbeck87} counterpart in IR color-color plots, \citet{Weight+94} identified V3645 Sgr with a likely giant counterpart. 
However, \citet{Surina14} observes a spectrum of this star, and find no evidence of either a giant or a white dwarf; they observe a star of spectral type K1-M1, probable luminosity class V or IV, and no evidence of emission lines. 

We obtained a SOAR spectrum (same setup as for V1310~Sgr on 2021 Jul 16.1) of the source suggested by \citet{Sarajedini84}, and again observe a spectrum consistent with a cool star and no evidence of emission lines.

We conclude that the counterpart to V3645~Sgr is unknown; it is likely a fainter source corresponding to a typical CV with a dwarf donor. However, as we are being conservative, we can not exclude the speculative possibility that the Sarajedini counterpart is associated with V3645~Sgr, is located at great distance, and marks a giant remnant of a CV merger.

\subsection{V5558 Sgr (2007)}

\begin{figure}
\includegraphics[angle=90,width=0.8\linewidth]{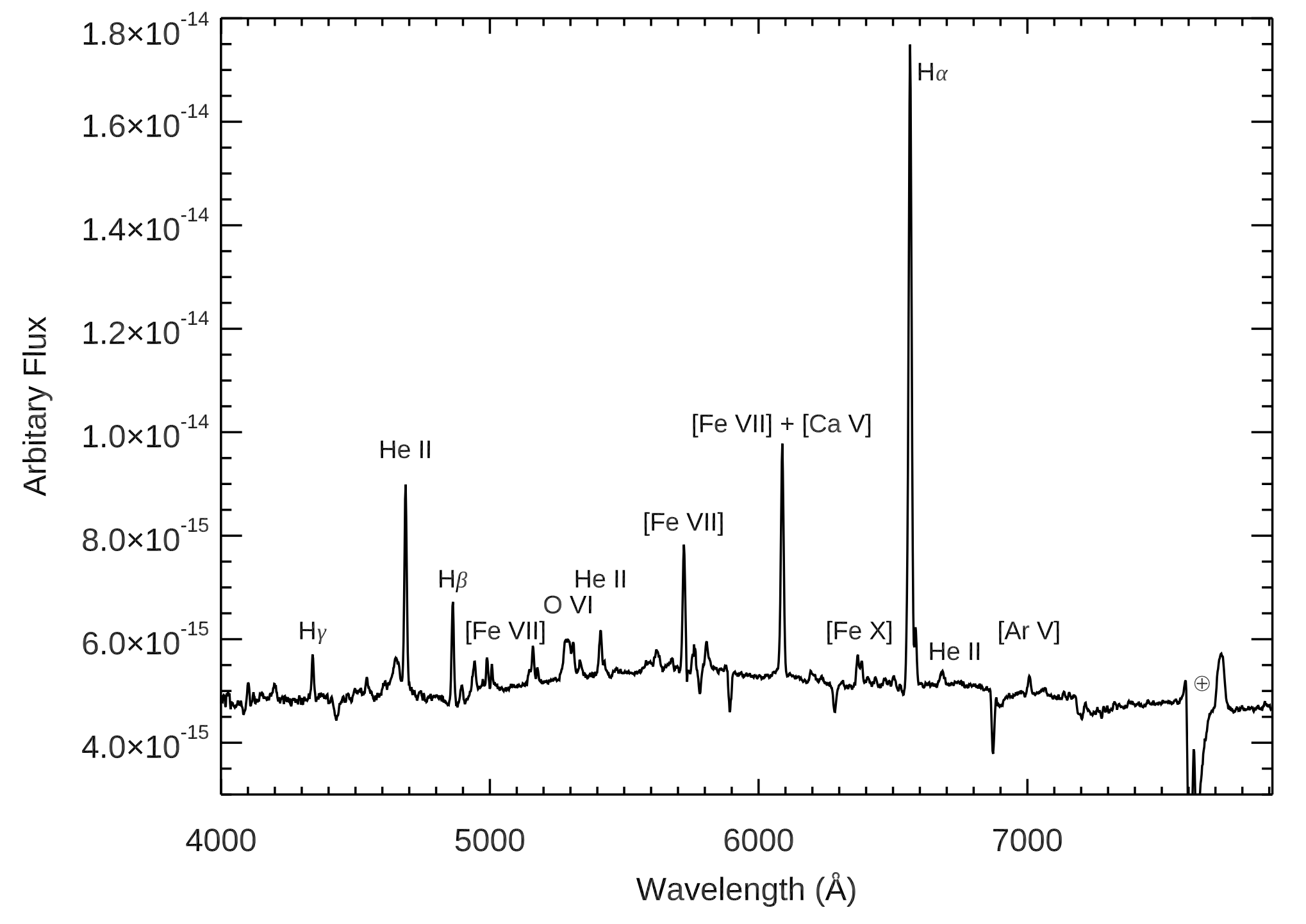}
\caption{SOAR Spectrum of V5558 Sgr showing high-ionization emission lines and a blue/flat continuum. A relative flux calibration has been applied.} 
\label{fig:v5558sgr}
\end{figure}

V5558 Sgr erupted relatively recently in 2007, and there is little in the literature about its post-eruption state.  Fourteen years after nova eruption, it remains bright, at $g \approx 15$ mag in 2021 observations with the Zwicky Transient Facility \citep{Masci+19}. Using the same setup as for V1310 Sgr, we obtained a SOAR spectrum on 2021 Jul 6.1. The spectrum shows emission lines superimposed on a  flat/blue continuum (Figure \ref{fig:v5558sgr}). The emission lines are of high-ionization species like HeII, FeVII, and FeX, as expected for a nova remnant if the white dwarf is still burning H on its surface \citep[e.g.,][]{Schwarz+11}. Meanwhile, the continuum suggests the presence of a disk, and the lack of absorption lines (apart from telluric bands or features associated with the interstellar medium) lead us to conclude that any companion star present is low luminosity and unevolved.


\end{document}